\documentclass[aps,preprintnumbers,superscriptaddress,nofootinbib]{revtex4}
\usepackage[dvips]{graphicx}
\usepackage{bm,latexsym,amsmath,amssymb,amsfonts,mathrsfs}
\usepackage{color}
\input{colordvi.tex}

\usepackage{dirtytalk}
\usepackage{mathtools,mathbbol}
\usepackage{graphicx,hyperref,xcolor}
\usepackage{lipsum}
\usepackage{float}
 \usepackage{slashed}
\def\Hc{\mathcal{H}}
\def\Pc{\mathcal{P}}

\def\Tc{\mathcal{T}}
\def\Ts{\text{T}}

\def\Fc{\mathcal{F}}
\def\Ec{\mathcal{E}}
\def\Lc{\mathcal{L}}
\def\Oc{\mathcal{O}}
\newcommand{\LF}{\left(}
\newcommand{\RF}{\right)}
\newcommand{\LT}{\left[}
\newcommand{\RT}{\right]}
\newcommand{\pd}{\partial}

\usepackage[normalem]{ulem}

\begin{document}


\title{CMB parity asymmetry from unitary quantum gravitational physics}

\author{Enrique Gazta\~naga}
\affiliation{Institute of Cosmology \& Gravitation,
	University of Portsmouth,
	Dennis Sciama Building, Burnaby Road,
	Portsmouth, PO1 3FX, United Kingdom}
\affiliation{Institute of Space Sciences (ICE, CSIC), 08193 Barcelona, Spain}
\affiliation{Institut d\'~Estudis Espacials de Catalunya (IEEC), 08034 Barcelona, Spain}

\author{K. Sravan Kumar}
\affiliation{Institute of Cosmology \& Gravitation,
	University of Portsmouth,
	Dennis Sciama Building, Burnaby Road,
	Portsmouth, PO1 3FX, United Kingdom}

\begin{abstract}

Longstanding anomalies in the Cosmic Microwave Background (CMB), including the low quadrupole moment and hemispherical power asymmetry, have recently been linked to an underlying parity asymmetry. We show here how this parity asymmetry naturally arises within a quantum framework that explicitly incorporates the construction of a geometric quantum vacuum based on parity ($\Pc$) and time-reversal ($\Tc$) transformations. This elegant framework restores unitarity in quantum field theory in curved spacetime (QFTCS). 
When applied to inflationary quantum fluctuations, this unitary QFTCS formalism predicts parity asymmetry as a natural consequence of cosmic expansion, which inherently breaks time-reversal symmetry. Observational data strongly favor this unitary QFTCS approach, with a Bayes factor, the ratio of marginal likelihoods associated with the model given the data $p\LF M\vert D \RF$, exceeding 650 times that of predictions from the standard inflationary framework. 
This Bayesian approach contrasts with the standard practice in the CMB community,
which evaluates $p\LF D\vert M \RF$, the likelihood of the data under the model, which undermines the importance of low-$\ell$ physics. Our results, for the first time, provide compelling evidence for the quantum gravitational origins of CMB parity asymmetry on large scales.

\end{abstract}

\maketitle

\section{Introduction}
\label{S:1}

Understanding and testing the interplay between gravity and quantum mechanics (QM) is the most important key to unraveling the new physics beyond the standard model of particle physics and Einstein's general relativity (GR). The merger of GR and QM is an essential physics that can be unlocked at the length scales of gravitational horizons. For this, we perhaps need an in-depth understanding of how one can resolve an age-old conundrum in the concept of time between the GR and QM.  The success of Quantum field theory (QFT) in Minkowski spacetime (\cite{Coleman:2018mew}) already teaches us several lessons in this regard. Indeed, QFT is the culmination of special relativity (SR) and QM, where the concept of time already enters into the conflict. In SR, the space and time are on equal footing, but in QM, spatial coordinates are elevated to operators while time is just a parameter due to its anti-unitary nature under the discrete transformation ($\Tc: t\to -t$). The SR and QM are merged by imposing the commutativity of operators corresponding to space-like distances. The underlying reason is that SR forbids any communication happening in space-like distances, and we respect it and achieve it by formulating new mathematical rules via second quantization. This is exactly the lesson one must keep in mind to understand quantum fields in curved spacetime. That is, we take lessons from classical physics and comply with them according to the rules of quantum physics. The quantum field theory in curved spacetime (QFTCS) is the first step in seeing GR and QM acting together, besides the renormalizable ultra-violet complete quantum gravity that may be required at the Planck scales. Furthermore, QFTCS is a necessity in the context of understanding inflationary quantum fluctuations (\cite{Mukhanov:1981xt,Sasaki:1986hm,Albrecht:1992kf}). The longstanding problem of QFTCS has been the loss of unitarity and the information paradox, which was first formulated in the context of black holes by Hawking's seminal works \cite{Hawking:1974rv,Hawking:1975vcx} and later extended to de Sitter space in the paper of Gibbons and Hawking \cite{Gibbons:1977mu}. The problem of unitarity emerges with the way quantization is usually done in curved spacetime, leading to the phenomenon of pure states evolving into mixed states. It means, an observer bounded by a spacetime horizon loses access to the states beyond the horizon. 
This problem first identified by Schr\"{o}dinger in 1956 in the context of de Sitter space \cite{Schrodinger1956}. Despite some attempts in the setups of de Sitter holography  \cite{Parikh:2002py,Parikh:2004wh} which attempted to save unitarity, multiple investigations in recent decades have largely focused on accepting unitarity loss and proposing entangled pairs of universes and parallel spacetimes or restoring unitarity from ad-hoc yet unknown frameworks of quantum gravity, See \cite{Almheiri:2020cfm,Hartman:2020khs,Shaghoulian:2022fop,Balasubramanian:2001rb,Balasubramanian:2021wgd,Shaghoulian:2021cef,Giddings:2022jda} and references therein. Motivated from 't Hooft, Schr\"{o}dinger and Einstein-Rosen's seminal works \cite{tHooft:2016qoo,Schrodinger1956,Einstein:1935tc} which insisted on the role of parity and time reversal operations, the recent proposal of direct-sum quantum field theory (DQFT) restores the unitarity in curved spacetime \cite{Kumar:2023ctp,Kumar:2023hbj,GKM,Kumar:2024ahu,Kumar:2024oxf} along with the potential resolution to the information-loss paradox.    

In this paper, we address crucial questions related to the meaning of time reversal and parity inversions in the context of inflationary quantum fluctuations and uncover their true nature in the cosmic microwave background (CMB). We explain how DQFT \cite{Kumar:2023ctp,Kumar:2023hbj,GKM}) is a necessity to solve the unitarity problem of QFTCS. The question of unitarity in QFTCS precedes the question of a completely renormalizable quantum gravity theory of Planck scales. In contrast to the widespread schools of thought that argue unitarity in QFTCS needs to be given up within GR and QM, we propose a new fundamental understanding of this issue and a new observational test using the parity asymmetry in the CMB. We find that the theory of inflationary quantum fluctuations with DQFT (which we call shortly DSI, which stands for "Direct-Sum inflation") fits up to 650 times better than the Standard Inflation (SI). This paper ultimately presents the picture of how a new theory of QFTCS emerged from fundamental questions of quantum gravitational physics that can potentially explain the observed CMB (large-scale) anomalies. The new results in this paper are further complemented by our more extended companion paper (\cite{Gaztanaga:2024vtr}). 

The paper is organized as follows: In Sec .~\ref {sec:dqft} we summarize the formulation of DQFT, which is built on a new understanding of quantum theory with two arrows of time along with parity-based geometric superselection rules. In Sec.~\ref{sec:dsdqft}, we discuss the construction of DQFT in de Sitter spacetime and demonstrate how we achieve unitarity in this setup. In Sec .~\ref {sec:cmbparity}, we present the application of DQFT to inflationary quantum fluctuations (i.e., DSI) and provide the strong statistical evidence for it over the standard framework of inflation (SI) using the latest CMB data from Planck. In Sec .~\ref {sec:concl} we conclude the important elements of our investigation, which expose the role of gravity and quantum mechanics in the early universe and the fresh perspective we bring to the subject with compelling observational evidence. In Appendices, \ref{app:QHO}, \ref{ap:KG} and \ref{ap:dS}, we provide further essential details of {direct-sum QM}, DQFT in Miknowski and de Sitter spacetimes.

\section{DQFT, in a nutshell ($\hbar=c=1$)}
\label{sec:dqft}

DQFT formulation emerges from a simple reconstruction of a quantum state or a quantum field operator through the discrete (a)symmetries (parity $\Pc$ and time reversal $\Tc$) of the background spacetime. Here, we summarize the conceptual understanding of DQFT and direct the reader to \cite{Gaztanaga:2024vtr,Kumar:2023ctp,Kumar:2023hbj,GKM,Kumar:2022zff,Kumar:2024ahu,Kumar:2024oxf} for more details. Construction of QFT in Minkowski spacetime requires the definition of a positive energy state and the arrow of time \cite{Donoghue:2019ecz}. Here, we build quantum theory with two arrows of time using the direct-sum Schr\"{o}dinger equation
	\begin{equation}
		i\frac{\pd\vert \Psi\rangle}{\pd t_p} = \begin{pmatrix}
			\hat{\mathbb{H}}_+ & 0 \\ 
			0 & -\hat{\mathbb{H}}_-
		\end{pmatrix} \vert \Psi\rangle 
\label{eq:Schr}
 \end{equation}
	where 
    \begin{equation}
         \hat{\mathbb{H}}\LF \hat x,\,\hat p \RF= \hat{\mathbb{H}}_+\LF \hat x_+,\, \hat p_+ \RF\oplus \hat{\mathbb{H}}_-\LF \hat x_-,\, \hat p_- \RF = \begin{pmatrix}
	    \mathbb{H}_+ && 0 \\ 
        0 && \mathbb{H}_-
	\end{pmatrix}
    \end{equation}
 it is the time-independent Hamiltonian operator (which is Hermitian) {which we assume here to be} $\Pc\Tc$ symmetric for simplicity.
 \footnote{{It is important to note that our construction is not related to $\Pc\Tc$ symmetric QM where systems with non-Hermitian Hamiltonians are studied \cite{Bender:1998gh}. In contrast, we work here with Hermitian Hamiltonian operators.}} Here, the position operator is split into direct-sum of two components ($\hat x_\pm$) as 
    \begin{equation}
    \hat x = \hat x_+\oplus \hat x_- = \begin{pmatrix}
        \hat x_+ && 0 \\
        0 && \hat x_-
    \end{pmatrix}
    \end{equation}
which have parity-conjugate eigenvalues when they act on the states $\vert \Psi_\pm\rangle$. Similarly, the momentum operator is split as 
\begin{equation}
    \hat p =\hat p_+\oplus \hat p_- = \begin{pmatrix}
        \hat p_+ && 0 \\
        0 && \hat p_-
    \end{pmatrix}
\end{equation}
with $\hat p_\pm$ as the momentum operators associated with the states $\vert \Psi_\pm\rangle $ that correspond to Hilbert spaces representing parity conjugate regions of physical space as explained further below.  
    The quantum state $\vert \Psi\rangle $ in Eq.~\ref{eq:Schr} is expressed as the direct-sum of two components 
\begin{eqnarray}
		\vert \Psi \rangle & = &  \frac{1}{\sqrt{2}} \LF \vert \Psi_{+}\rangle \oplus \vert \Psi_{-} \rangle \RF = \frac{1}{\sqrt{2}} \begin{pmatrix}
			\vert \Psi_{+}\rangle \\ \vert \Psi_{-}\rangle
		\end{pmatrix}
		\label{psid}
 \\ \int_{-\infty}^\infty \langle \Psi\vert\Psi \rangle dx & = & \frac{1}{2} \int_{-\infty}^\infty\Big[\langle \Psi_+\vert\Psi_+ \rangle  +  \langle \Psi_-\vert\Psi_- \rangle \Big] dx=1\,.
\nonumber
\label{disumwvf}
\end{eqnarray}
where $\vert \Psi_+\rangle,\,\vert \Psi_-\rangle$ are the two components of the same state at parity conjugate points. The component $\vert \Psi_+\rangle$
is a positive energy state ($\Ec>0$) that evolves forward in time as $\vert \Psi_+\rangle_{t_p}= e^{-i\Ec t_p}\vert \Psi_+\rangle_{0}$ where the parametric time ($t_p$) arrow $t_p: -\infty \to \infty$ while $\vert \Psi_-\rangle$ can also be seen as a positive energy state that evolves back in time as $\vert \Psi_-\rangle_{t_p}= e^{i\Ec t_p}\vert \Psi_-\rangle_{0}$ with the parametric time convention $t_p: +\infty \to -\infty$. One can notice here the purely antiunitary character of time associated with the replacement $i\to -i$ (\cite{Donoghue:2019ecz}). Note that the minus sign in front of $\mathbb{H}_-$ in Eq.~\ref{eq:Schr} represents the opposite arrow of time. Thus,
the direct-sum Schr\"{o}dinger equation does not have any arrow of time associated with it compared to the conventional Schr\"{o}dinger equation. To be explicit, we can rewrite Eq.~\ref{eq:Schr} as
\begin{equation}
    \begin{pmatrix}
        i\frac{\pd}{\pd_{t_p}}\vert \Psi_+\rangle \\ 
       -i \frac{\pd}{\pd_{t_p}} \vert \Psi_-\rangle \end{pmatrix}  = \begin{pmatrix}
	    \hat{\mathbb{H}}_+ && 0 \\ 
        0 && \hat{\mathbb{H}}_-
	\end{pmatrix} \begin{pmatrix}
	    \vert \Psi_+ \rangle \\ 
        \vert \Psi_-\rangle
	\end{pmatrix}
\end{equation}
The direct-sum splitting of the quantum state as in Eq.~\ref{psid} implies the Hilbert space ($\Hc$) splitting into a direct-sum of the two ($\Hc_\pm$) corresponding to parity conjugate points in position space:
 $   \Hc = \Hc_+ \oplus \Hc_-$.
Here, the Hilbert spaces $\Hc_\pm$ are called the geometric superselection sectors (\cite{GKM,Kumar:2023ctp}).
The position and momentum commutation relations now become double associated with parity conjugate position and momentum space operators denoted by subscripts $+,\,-$, and their relations are
\begin{equation}
\Big[ \hat x,\, \hat p \Big] = i \implies \begin{cases}
    \Big[\hat{x}_+,\,\hat{p}_+\Big]  = - \Big[\hat{x}_-,\,\hat{p}_-\Big]= i\, \\ \nonumber
    
  [\hat{x}_+,\,\hat{x}_-] =[\hat{p}_+,\,\hat{p}_-] = [\hat x_{\pm},\,\hat p_{\mp}]=0
\end{cases}, \quad  \hat{p}_\pm=\mp i \frac{\pd}{\pd x_\pm},
\quad 
{\begin{aligned}
x_+ & = x \in (0,\infty] \\ 
x_- & = x \in [-\infty, 0) \\ 
\end{aligned}}
\end{equation}
{The concept of wavefunction in the direct-sum QM becomes 
\begin{equation}
\Psi(x) = \frac{1}{\sqrt{2}} \begin{pmatrix}
       \langle x_+\vert \quad \langle x_- \vert 
    \end{pmatrix}\begin{pmatrix}
        \vert \Psi_+\rangle_0 e^{-i\Ec t_p} \\
        \vert\Psi_-\rangle_0 e^{i\Ec t_p}  
    \end{pmatrix} \implies \begin{cases}
        \frac{1}{\sqrt{2}}\Psi_+\LF x_+ \RF e^{-i\Ec t_p},\quad x_+ = x\gtrsim 0 \\  \frac{1}{\sqrt{2}}\Psi_-\LF x_- \RF e^{i\Ec t_p},\quad x_- = x\lesssim 0\,.
    \end{cases}
    \label{eq:wavefunction}
\end{equation}}
This framework makes the wavefunction of the quantum system explicitly $\Pc\Tc$ symmetric that resonates with the assumed $\Pc\Tc $ symmetry of the physical system, i.e. $[\mathbb{H},\,\Pc\Tc]=0$ and the results of standard quantum mechanics remain the same as shown with the example of harmonic oscillator worked out in \cite{Gaztanaga:2024vtr} (see also further discussions and Fig.~1 in \cite{Kumar:2023ctp,Kumar:2024ahu}). 
The DQFT in Minkowski spacetime ($ds^2= -dt_m^2+d\textbf{x}^2$ ($t_m$ being the time coordinate) which is $\Pc\Tc$ symmetric i.e., $t_m\to -t_m$,\,$\textbf{x}\to -\textbf{x}$) is just built on the direct-sum Schr\"{o}dinger equation Eq.~\ref{eq:Schr}, following how the standard QFT is built on the definition of positive energy state and the causality condition (i.e., commutativity of operators for spacelike distances). See Appendix.~\ref{app:QHO} and \ref{ap:KG} for more details.


\section{DQFT in de Sitter \& Unitarity ($G=\hbar=c=1$)}
\label{sec:dsdqft}

The de Sitter (dS) spacetime is a perfect testing ground for exploring the theory of QFTCS because of its maximally symmetric aspect, and it is also relevant for unlocking the nature of inflationary quantum fluctuations. 
Furthermore, dS spacetime has the closest resemblance to black holes, where the most important conundrums in quantum physics emerge, such as the loss of unitarity and the information-loss paradox \cite{Gibbons:1977mu}. 
We start with the dS metric in the flat Friedman-Lema\^itre-Robertson-Walker (FLRW) coordinates 
\begin{equation}
    ds^2 = -dt^2+e^{2Ht}d\textbf{x}^2 = \frac{1}{H^2\tau^2}\LF -d\tau^2+d\textbf{x}^2 \RF \,, 
    \label{dSmetric}
\end{equation}
where the scale factor $a=e^{Ht}$ is the clock that determines the expansion of the Universe, $H=\frac{\dot{a}}{a}$ is the Hubble parameter and $\tau = \int \frac{dt}{a} = -\frac{1}{aH}$ is the conformal time for $a=e^{Ht}$. If one has to place quantum theory in dS spacetime Eq.~\ref{dSmetric}, the first and foremost things to be understood are the parity and time reversal operations. The usual conception of understanding the expansion of the Universe is to restrict ourselves to $\tau<0$. But the metric carries a symmetry 
\begin{equation}
    \Tc: \tau\to -\tau,\quad \Pc:\textbf{x}\to -\textbf{x}
    \label{sym}
\end{equation}
Restricting to $\tau<0$ before quantization of a classical field (as is usually followed in standard QFTCS) means throwing away the symmetry in Eq.~\ref{sym} by hand. The expanding Universe can be described in two ways:
\begin{equation}
\begin{aligned}
 & \implies \begin{cases}
		t: -\infty \to +\infty,\quad  H>0 & (\tau: -\infty \to 0)\\ 
		t: +\infty \to -\infty,\quad  H<0 & (\tau: \infty \to 0)
	\end{cases}
 \end{aligned}
\label{expcon}
\end{equation}
Notice in particular that in Eq.~\ref{expcon}, reversing the arrow of time together with the sign of the Hubble parameter represents the same expanding universe. This means that the scale factor is the clock that determines the expansion of the universe.
This would explain why in the Wheeler-de Witt equation the cosmic time does not explicitly appear; rather, the scale factor is the clock that runs with the expanding universe \cite{Kiefer:2007ria}. Note that $H\to -H$ does not change any properties of the dS space at all. For example, curvature invariants such as the Ricci scalar $R=12H^2$ remain the same under a discrete sign flip of H. 
Eq.~\ref{expcon} illustrates two possible time realizations for an expanding universe. By convention, choosing $\tau>0$ or $\tau<0$ results in loss of information beyond the horizon.  
In his famous monograph of 1956, Schrodinger \cite{Schrodinger1956} rejected the idea of two universes because it would lead to a conundrum of ignorance of information beyond the horizon, which in modern language means pure states evolving into mixed states which leads the violation of unitarity and information-loss (\cite{Gibbons:1977mu,Parikh:2002py}).

\begin{figure}
    \centering
    \includegraphics[width=0.8\linewidth]{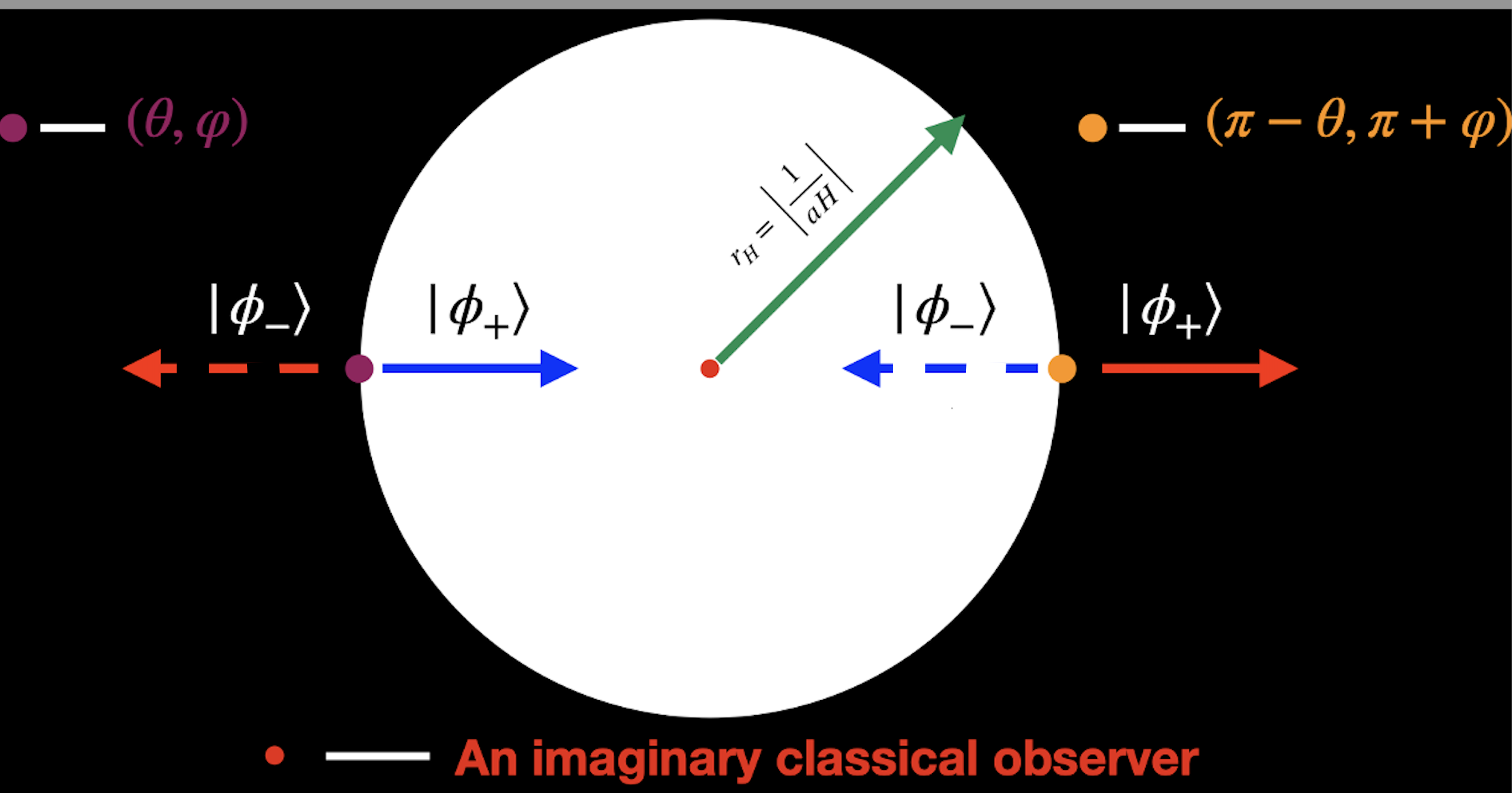}
    \caption{This figure depicts the concept of unitarity that can be novelly achieved in de Sitter spacetime (in flat FLRW coordinates) with DQFT. Given the particle horizon ($r_H$) at any moment of de Sitter expansion, the above picture depicts how the DQFT formulation projects all information of the states outside the horizon back inside at the antipodal point of the physical space. This implies that all the information that any imaginary classical observer perceives is within the horizon and that pure states evolve into pure states. Blue arrows indicate accessible from inside the horizon, while dashed lines indicate that in the quasi-deSitter case the time evolution of $ \vert \phi_+\rangle$ and  $\vert \phi_-\rangle$ could be different, which is the origin of creating CMB parity asymmetry in the large scales.}
    \label{fig:unitarity}
\end{figure}

Similar to the Minkowski space (see Appendix.~\ref{ap:KG}), the quantum fields in the dS space are described through a direct-sum vacuum $\vert 0\rangle_{\rm dS} = \vert 0_+\rangle_{\rm dS}\oplus \vert 0_-\rangle_{\rm dS} $  corresponding to the direct-sum Fock space $\Fc_{\rm dS}=\Fc_{\rm dS+}\oplus \Fc_{\rm dS-}$. This implies that everywhere in dS space, a quantum field is tied by time forward and backward evolutions at the parity conjugate points, see Appendix.~\ref{ap:dS}. At any moment of dS expansion, the parity-conjugate points on the comoving horizon $r_H= \vert \frac{1}{aH}\vert$ are space-like separated for an imaginary classical observer as shown in Fig.~\ref{fig:unitarity}. 
Note that by DQFT construction, the direct-sum quantum states $\vert\phi_\pm\rangle = \hat{d}_{(\pm)\,\textbf{k}}^\dagger \vert 0_{\pm}\rangle_{\rm dS}$ that follow from Eq.~\ref{fiedDQFTdS} are the opposite time evolving positive energy states at the parity conjugate points and this holds for any imaginary classical observer. Furthermore, in DQFT, causality and locality are preserved too. A consequence of this is that any state that disappears beyond the horizon (say at $\LF \theta,\,\varphi \RF$) happens to be the complementary direct-sum (component) state that reappears at the antipodal point (at $\LF \pi-\theta,\,\pi+\varphi \RF$). This is because all the imaginary classical observers, including those on the horizons of each other, should observe the same direct-sum quantum state $\vert \phi_{+}\rangle \oplus \vert \phi_{-} \rangle $.

\begin{figure}
    \centering
    \includegraphics[width=.8\linewidth]{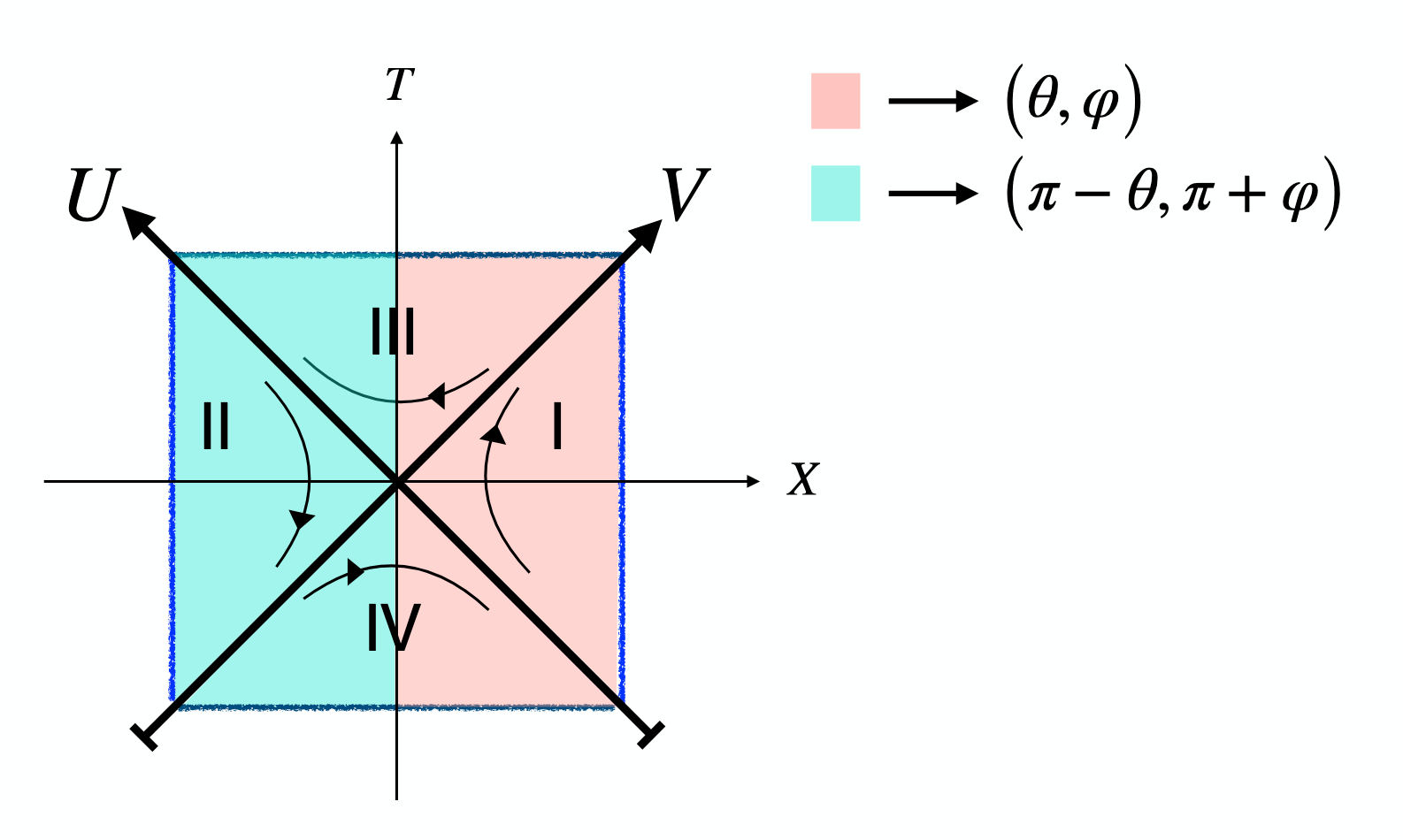}
    \caption{This picture depicts the (quantum) conformal diagram of static dS spacetime Eq.\ref{statdS} with four regions that are related by discrete transformations and the lines with arrows indicate the arrows of time in each region. Here a quantum field operator is written as direct-sum of four components $\hat \phi= \frac{1}{\sqrt{2}}\LF \hat \phi_I\oplus \hat\phi_{II}\RF \oplus \frac{1}{\sqrt{2}}\LF \hat\phi_{III}\oplus \hat \phi_{IV}  \RF  $ corresponding to the regions I to IV. This is not a Penrose diagram because each point represents half of the $S_2$ sphere, in contrast to the usual Penrose diagram. This diagram is to represent quantum physics in dS space rather than classical physics, for which the usual Penrose diagram is used.  }
    \label{fig:NPD}
\end{figure}

This means the horizon acts as a $\Pc\Tc$ mirror which maps every information outside the horizon to the states inside the horizon. 
To see this, consider a pure state:
\begin{equation}
    \vert \psi \rangle = \sum_{m} d_{m} \vert \phi_{m} \rangle \,,
\end{equation}
where $ \vert \phi_{m} \rangle$ are the eigenvectors, i.e. for any observable (an operator) $\hat{O}$ we have: $\hat{O}  \vert \phi_{m} \rangle =\lambda_m  \vert \phi_{m} \rangle$ where $\lambda_m$ are the corresponding  eigenvalues. Each  $\lambda_m$  can be observed  in $ \vert \psi \rangle $ with a probability: $P(\lambda_m)=|d_m|^2$. Consider now the (de Sitter) Hubble horizon $r_H= |\frac{1}{aH}|$ around an observer (as illustrated in Fig.~\ref{fig:unitarity}).
 In the standard approach, information is lost (a pure state becomes a mix state) because communication can not happen for space-like distances (i.e. outside the horizon) as was shown by Gibbons and Hawking in 1977 \cite{Gibbons:1977mu}, in analogy to what happens in Hawking radiation and the black hole information loss paradox \cite{Hawking:1974rv,Kumar:2023hbj}. This also implies the loss of unitarity in QFT on spacetimes with horizons.
In the DQFT approach, we have:
\begin{equation}
    \vert \psi\rangle = 
      \frac{1}{\sqrt{2}}\begin{pmatrix}
        \sum d_{m}^+ \vert \phi_m^+ \rangle \\ 
        \sum d_{m}^- \vert \phi_m^- \rangle 
    \end{pmatrix} 
     \equiv \frac{1}{\sqrt{2}}\begin{pmatrix}
        \vert \psi^+\rangle \\ 
        \vert \psi^-\rangle 
    \end{pmatrix}
    \label{puredS}
\end{equation}
where  $ \vert \phi_{m}^\pm \rangle$ are the eigenvectors corresponding to each  $\Pc\Tc$ separated  Hilbert space. So that the observable probabilities and operators become:
	\begin{equation}
    P(\lambda_m^\pm)=|d_m^\pm|^2 \quad \text{and} \quad 
		\hat{O} = \begin{pmatrix}
			\hat{O}_+ & 0 \\ 
			0 & \hat{O}_-
		\end{pmatrix} 
 \end{equation}
Information is not lost in this case because there is not entanglement across the horizon. 
To see this more clearly,
consider the case where $\vert \psi \rangle$ is a pure entangled system 
$\vert \psi_{AB} \rangle = \vert \psi_{A} \rangle \otimes  \vert \psi_{B} \rangle$:
\begin{equation}
    \vert \psi_{AB} \rangle = \sum_{m,n} d_{mn} \vert \phi_{Am} \rangle \otimes  \vert \phi_{Bn}\rangle \,,
\end{equation}
where $\otimes $ represents the tensor product of two Hilbert spaces $A$ and $B$ which are entangled (inseparable or correlated): $d_{mn} \neq d_m d_n$ with:  $\vert \phi_A \rangle = \sum_m d_m\vert \phi_{Am}\rangle $ and $\vert \phi_B \rangle = \sum_n d_n\vert \phi_{Bn}\rangle $. For example, a pair of particles (across the horizon).
In the DQFT approach, we have:
\begin{equation}
    \vert \psi_{AB}\rangle
=    \frac{1}{\sqrt{2}}\begin{pmatrix}
        \sum d_{mn}^+ \vert \phi_{Am}^+ \rangle \otimes  \vert \phi_{Bn}^+\rangle \\ 
        \sum d_{mn}^- \vert \phi_{Am}^- \rangle \otimes  \vert \phi_{Bn}^-\rangle
    \end{pmatrix} 
     \equiv \frac{1}{\sqrt{2}}\begin{pmatrix}
        \vert \psi_{AB}^+\rangle \\ 
        \vert \psi_{AB}^-\rangle 
    \end{pmatrix}
    \label{puredS}
\end{equation}
where $ \vert \phi_{Am}^\pm \rangle$ and  $ \vert \phi_{Bn}^\pm \rangle$ are the eigenvectors corresponding to each  $\Pc\Tc$ separated  Hilbert spaces, which remain within the horizon because no information is exchanged across $\Pc\Tc$ separated spaces. The fact that there are no cross terms in Eq.~\ref{puredS} follows from the definition of superselection sector Hilbert spaces, according to which one cannot form any superposition of states from the superselection sectors (See \cite{Conway,Giulini:2007fn,nlab:superselection_theory} for more details on superselection sectors and direct-sum Hilbert spaces and the associated tensor products). This is why in Eq.~\ref{puredS} we do not have any cross-products between parity conjugate states. All information about the entanglement pairs is mapped within the horizon; thus, the pure state will remain pure. In other words, DQFT forbids any entanglement beyond the gravitational horizons. 
DQFT constructs a direct-sum Hilbert space with the interior and exterior separated by a geometric superselection rule (which is similar to the concept of superselection sectors in QFT   \cite{Wick:1952nb,nlab:superselection_theory}).
Thus, there would be an observer complementarity (see Fig.~3 in \cite{Kumar:2023ctp}).
We can also understand the same from the point of view of dS space in static coordinates, which reads as 
\begin{equation}
\begin{aligned}
    ds^2 & = -\LF 1-H^2r^2 \RF dt_s^2 + \frac{1}{\LF 1-H^2r^2 \RF}dr^2 + r^2d\Omega^2 \\ 
    &  = \frac{1}{H^2\LF 1-UV \RF^2} \LF -4dUdV+ \LF 1+UV \RF^2 d\Omega^2 \RF
    \end{aligned}
    \label{statdS}
\end{equation}
where $r= \big\vert \frac{1}{H}\big\vert \frac{1+UV}{1-UV}$ and $\frac{V}{U} = -e^{2Ht_s}$. The relation between the flat FLRW Eq.\ref{dSmetric} and Eq.\ref{statdS} is just a simple coordinate change (\cite{Lanczos,Hartman:2017}).
A very practiced view of flat dS spacetime in Eq.~\ref{dSmetric} is that it covers only half of the dS space. However, we must keep in mind the symmetry in Eq.~\ref{expcon}, which implies that Eq.~\ref{dSmetric} can cover the entire dS space. Thus, the maximally symmetric nature of the dS spacetime sets no difference whether it is expressed in flat, closed, open FLRW coordinates or the static coordinates (\cite{Mukhanov:2007zz,Kumar:2023ctp}). The use of mapping flat coordinates to static coordinates here is so that we can realize how unitarity can be achieved with DQFT, whereas it is lost in standard QFT, as was shown by 
Gibbons and Hawking in 1977 \cite{Gibbons:1977mu}.
In Fig.~\ref{fig:NPD}, we show unitary quantum physics in the static dS space through the (quantum) conformal
diagram, where we see the three-dimensional space divided by parity for the interior and exterior of the cosmological event horizon. We can picture here how none of the observers loses any information beyond the horizon. If any state leaves the horizon, it is only the (complimentary) direct-sum component of what is inside the horizon. This means that any observer can perfectly reconstruct the information about what is beyond the horizon. According to DQFT, every density matrix of a maximally entangled pure state is split into direct-sum of 4 components 
\begin{equation}
    \rho= \LF \rho_{I}\oplus \rho_{II}\RF \oplus \LF \rho_{III}\oplus \rho_{IV} \RF
\end{equation}
Each density matrix $\rho_{n},\,n=I,II,III, IV$ is defined by a pure state component in the corresponding (geometric) superselection sector Hilbert space $\Hc_n$. Since the Von Neumann entropy of each component of the density matrix vanishes, $S_n = -Tr\rho_n\ln\rho_n =0$, any observer in regions I, II, III, IV (see Fig.~\ref{fig:NPD}) would witness pure states evolving into pure states. Thus, with DQFT, we can have both unitarity and observer complementarity in dS (\cite{Kumar:2023ctp}). Furthermore, this DQFT construction is also studied in the context of achieving unitarity in Rindler spacetime \cite{Kumar:2022zff}.

According to DQFT, any state is spread across the horizon by distinct components in geometric superselection sectors. This implies that any information of the entangled state is shared by the sectorial Hilbert space, leading to pure states evolving into pure states. To be precise, the Hilbert spaces of the interior and exterior states cannot be connected by direct product ($\otimes$) but rather should be related by direct-sum ($\oplus$). 
This has to be like this because beyond-the-horizon time is not the same, so one cannot do the same quantum mechanics everywhere. In other words, the gravitational horizons are special; they separate different spacetimes by a boundary. Thus, they cannot be treated as a standard null surface, where quantum states can be entangled across. Furthermore, we stress that no entanglement across the horizon does not lead to firewalls, which happens in standard QFTCS but not in direct-sum QFTCS. For further analogous discussion on unitarity and no information loss in the context of the Schwarzschild black hole, see \cite{Kumar:2023hbj}. In a nutshell, DQFT offers a new understanding of dS space where any observer's physics of the universe is complete, and no information is lost outside any observer's universe. This is exactly what Schr\"{o}dinger envisioned in 1956 for a quantum theory to achieve \cite{Schrodinger1956}. 
The formalism of DQFT has certain similarities with 'tHooft's recent ideas on black holes \cite{tHooft:2016qoo}. However, there are some crucial differences, one of them is the construction of the geometric superselection rule that restructures a single quantum state in sectorial Hilbert spaces that represent parity conjugate regions of physical space. 'tHooft on the other hand considers anti-podal identification of time-reversal regions of spacetime with a slightly different interpretation of parity. The idea of antipodal identification was later given up by 't Hooft because of the possibility of closed timeline curves when dynamics of spacetime are involved, so a new idea of quantum clones is proposed \cite{tHooft:2022bgo}. Our framework significantly differs from this latter proposal; we speak about $\Pc\Tc$ based mirror symmetries within the description of quantum fields in the same manifold rather than creating clones in different manifolds. One can also feel some aesthetic similarities between our proposal and Boyle and Turok's CPT symmetric universes \cite{Boyle:2018tzc,Boyle:2018rgh} (coincidentally, it is a new understanding of the old proposal by Sakharov \cite{Sakharov:1967dj,Sakharov:1980kc}). The crucial difference in our framework is that we are speaking about parity and time reversal within the same expanding universe, in contrast with Boyle and Turok's pair of CPT-symmetric universes before and after the Big Bang.   
Furthermore,  DQFT echoes in parallel with the recent classical understanding of dS through the Black Hole Universe (BHU) proposal \cite{Gaztanaga2022,gaztanaga2023a,Gaztanaga:2025cun}.

\section{CMB parity asymmetry}

\label{sec:cmbparity}

Inflationary quantum fluctuations are the initial seeds for large-scale structure formation. It is important to understand their quantum nature to derive consistent CMB predictions. Inflation is by definition a quasi-de Sitter expansion (\cite{Starobinsky:1980te}) and this means that $\Pc\Tc$ in Eq.~\ref{expcon} is spontaneously broken by the presence of a non-perturbative scalar field in addition to the GR's tensor degree of freedom. Our focus here is on single-field inflation. The metric and scalar field matter quantum fluctuations are generated during inflation. The quantity that is a gauge-invariant combination of them is called the curvature perturbation ($\zeta$), and it is an effective (scalar) degree of freedom ($\phi$). It connects the quantum physics of inflation to CMB observations.  The perturbed action for curvature perturbation at the linear order is \cite{Gaztanaga:2024vtr,Baumann:2018muz}
\begin{equation}\label{scalar}
		\delta^{(2)}S_{s} = \frac{1}{2}\int d\tau d^3x a^2\frac{\dot{\bar\phi}^{ 2}}{H^2} \Bigg[ \zeta^{\prime 2} -\LF \pd_i\zeta \RF^2 \Bigg]
		\,\text{,}
	\end{equation}
where $H = \frac{\dot{a}}{a}$ which is nearly constant during inflation characterized by the smallness of slow-roll parameters $\epsilon=-\frac{\dot{H}}{H^2},\,\eta=\frac{\dot{\epsilon}}{H\epsilon}$ for $N=50-60$ number of e-foldings. In Eq.~\ref{scalar}, overbar indicates the background field, overdot indicates time derivative, $^\prime$ denotes derivative with respect to conformal time, and $\pd_i=\frac{\pd}{\pd x^i}$ implies spatial gradient. 
The canonical variable which is a redefinition of curvature perturbation that we eventually quantize is
	\begin{equation}
		v=a\zeta \dot{\bar\phi}/H
		\label{canvar}
	\end{equation}
In the framework of Direct-sum Inflation (DSI), the canonical variable is promoted to field operator that is split into direct-sum of two components \begin{equation}
		\begin{aligned}
			\hat{v} 
   & = \frac{1}{\sqrt{2}} \begin{pmatrix}
				\hat{v}_{+} \LF \tau,\,\textbf{x} \RF & 0 \\ 
				0 & \hat{v}_- \LF -\tau,\, -\textbf{x} \RF
			\end{pmatrix}
		\end{aligned}
		\label{fieldmat}
	\end{equation}
which can be expanded as 
\begin{equation}
	\begin{aligned}
& \hat{v}_{\pm }=   
\int \frac{ d^3k}{\LF 2\pi \RF^{3/2}} \Bigg[ c_{\LF\pm \RF\textbf{k}} {v}_{\pm,\,k} e^{\pm i\textbf{k}\cdot \textbf{x}} + c_{\LF\pm \RF\textbf{k}}^\dagger {v}_{\pm,\,k}^\ast e^{\mp i\textbf{k}\cdot \textbf{x}} \Bigg]
\end{aligned}
\label{vid}
\end{equation}
This would also mean promoting the curvature perturbation as an operator in the following way, which involves classical rescaling 
\begin{equation}
    \hat \zeta = \frac{H}{a\dot{\bar\phi}}\Bigg\vert_{\rm classical} \hat v 
\end{equation}
These quantum fluctuations evolve in a direct-sum vacuum defined by
\begin{equation}
		\vert 0\rangle_{\rm qdS} = \vert 0_+ \rangle_{\rm qdS} \oplus \vert 0_-\rangle_{\rm qdS} = \begin{pmatrix}
			\vert 0_+\rangle_{\rm qdS} \\ 
			\vert 0_-\rangle_{\rm qdS}
		\end{pmatrix},\quad c_{\pm \textbf{k}}\vert 0_\pm\rangle_{\rm qdS}=0\,.
		\label{qdSmat}
	\end{equation}
in which they evolve forward and backward in time at the parity conjugate points in physical space, governed by the mode functions $v_{\pm,\,k}$ are the solutions of 
\begin{equation}
	v_{\pm,\,k}^{\prime\prime}+ \LF k^2-\frac{{\nu}_s^{\LF \pm\RF 2}-\frac{1}{4}}{\tau^2} \RF v_{\pm,\,k}^2 =0,\quad \nu_s^{\pm} \approx \frac{3}{2}\pm\epsilon\pm\frac{\eta}{2}
	\label{MS-equation}
\end{equation}
 {Recall that the choice of mode functions determines the vacuum (See Appendix~\ref {ap:dS} for the discussion in the context of de Sitter)}.
The solutions of \eqref{MS-equation} are
\begin{equation}
	\begin{aligned}
	{v}_{\pm,\,k} &  \approx \sqrt{\frac{1}{2k}} e^{\mp ik\tau}\LF 1\mp\frac{i}{k\tau} \RF
	\pm \LF \epsilon+\frac{\eta}{2} \RF \frac{\sqrt{\pi}}{2\sqrt{k}} \sqrt{\mp k\tau} \frac{\pd H^{(1)}_{\nu_s^{\pm}}\LF \mp k\tau\RF}{\pd \nu_s^{\pm}}\Big\vert_{\nu_s^{\pm}=3/2}\equiv v^{\rm dS}_{\pm, k} \LF 1\pm \Delta v\RF
\label{new-vac1BD}
\end{aligned}
\end{equation}
where $H^{(1)}_{\nu_s}\LF z \RF$ is the Hankel functions of the first kind {and
\begin{equation}
    v_{\pm, k}^{\rm dS} = \sqrt{\frac{1}{2k}} e^{\mp ik\tau}\LF 1\mp\frac{i}{k\tau} \RF,\quad \Delta v = \LF \epsilon+\frac{\eta}{2} \RF \LT \frac{1}{H_{3/2}^{(1)} \LF  \frac{k}{k_\ast} \RF} \frac{\pd H^{(1)}_{\nu_s}\LF \frac{k}{k_\ast} \RF}{\pd\nu_s} \Bigg\vert_{\nu_s=\frac{3}{2}} \RT
    \label{modeqds}
\end{equation}}
The mode functions of curvature perturbation can be expressed as:
\begin{equation}
    \zeta_{\pm k} = \frac{H}{a\dot{\bar\phi}} v_{\pm k} = \zeta_{\pm\,k}^{\rm dS}\LF 1\pm \Delta v\RF   \quad ; \quad \zeta_{\pm\,k}^{\rm dS} \equiv \frac{H}{a\dot{\bar\phi}} v_{\pm,\,\textbf{k}}^{\rm dS} 
    \label{eq:vpm}
\end{equation}
We can see here how the extra quasi-de Sitter contribution $\pm\Delta v$ is anti-symmetric in parity conjugate regions of space.
In analogy with Eq.~\ref{expcon}, the time reversal operation (quantum mechanically) in an expanding quasi-dS Universe is\footnote{Our notation of any quantity $Q\to -Q$ means we replace the quantities by a negative sign of that quantity.} 
\begin{equation}
		\tau\to -\tau \implies (t, H, \epsilon,\eta) \to (-t,-H,-\epsilon, -\eta). 
  \label{dsisym}
	\end{equation}
We stress that the above flip of the sign of the slow-roll parameters should not be thought of from a classical point of view. Indeed, if one does literally $ t\to-t$ and $ H\to-H$, the definition of slow-roll parameters does not change sign. However, what we have to keep in mind here is that quantum mechanically, the concept of time is different from the classical point of view. Time is a parameter in quantum theory and also in quantum field theory \footnote{In quantum mechanics, coordinate time is not an operator but a parameter because of the fundamental reason that time-reflection is an anti-Unitary operation \cite{Brunetti:2009eq,Donoghue:2019ecz,Donoghue:2021meq,Roberts:2022xcj}. It is worth recalling that in the usual QFT in Minkowski spacetime, which is the unification of special relativity and quantum mechanics, is built on assuming an arrow of time $t_p: -\infty \to \infty$, definition of positive energy ($\mathcal{E}>0)$ state $\vert\Psi\rangle_t = e^{-i\mathcal{E} t_p}\vert \Psi\rangle_0$ and the operators to commute for space-like distances. DQFT brings a new construction of quantum theory, which gives a new understanding of quantum physics with both arrows of time.  } \cite{Kumar:2023ctp}. When it comes to gravity, we encounter the dynamical nature of spacetime, so when we speak about quantum fields in curved spacetime, not only time but also any additional quantities that define the dynamics of spacetime act as parameters quantum mechanically. When we study inflationary fluctuations, which are a combination of metric and matter degrees of freedom (indeed the curvature perturbation $\zeta = -\Psi+H\frac{\delta\phi}{\dot{\bar\phi}}$ is a combination of Bardeen potential $\Psi$ and the inflaton field fluctuation $\delta\phi$), what we do is linearized quantum gravity (See \cite{Martin:2004um} for more discussion) that is encapsulated by the following equation which also include tensor fluctuations\footnote{In this paper, we only address the parity asymmetry in two-point functions of scalar fluctuations (that leads to asymmetry in the even-odd angular power spectra that is significant on large scales) according to DSI model but the DSI also predicts parity asymmetry in the tensor two point correlations as explained in \cite{Kumar:2022zff}. } 
    \begin{equation}
        \delta\hat G_{\mu\nu} = \delta \hat T_{\mu\nu}
    \end{equation}
The above sign changes (once more interpreted quantum mechanically) in the slow-roll parameters, which break the time-reversal symmetry of de Sitter spacetime, accommodate a new vacuum structure in quasi-de Sitter spacetime that lead to corrections in the evolution of quantum fluctuations $\hat{v}_{(\pm)}$ in the parity-conjugate regions. In the case of exact de Sitter spacetime, the DQFT framework leads to equal correlations for field components in the $\Pc\Tc$ conjugate regions, but in the quasi-de Sitter we do get unequal correlations because the formulation of our quasi-de Sitter vacuum Eq.~\ref{qdSmat} leads to fluctuations evolving asymmetrically at parity conjugate points in physical space. Thus, it is a pure quantum (gravitational) effect, and it has no classical analog. One can interpret this result as inflation breaks spontaneously the $\Pc\Tc$ symmetry of de Sitter space, implying the quantum fluctuations to evolve asymmetrically at parity conjugate points (see Fig.~3 in \cite{Gaztanaga:2024vtr} or Fig.\ref{fig:unitarity} above). As we will show, this results in an asymmetry in the CMB temperature under spatial parity. Following the two (scalar) power spectra defined with respect to the two quasi-de Sitter vacua Eq.~\ref{qdSmat} that correspond to parity conjugate regions of physical space. From Eq.\ref{eq:vpm} we have: 
\begin{equation}
    \Pc_{\zeta\pm}  = \int d^3r e^{-i\textbf{k}.\textbf{r}}	{}_{\rm qdS}\langle 0_{\pm}\vert  \hat \zeta_{\pm}\LF \pm\tau,\,\pm \textbf{x} \RF  \hat \zeta_{\pm}\LF \pm \tau,\, \pm\textbf{x}^\prime \RF\vert 0_\pm\rangle_{\rm qdS} \approx P_\zeta\LF 1\pm \Delta \Pc_v\RF
    \label{eq:pwdsi}
\end{equation} 
where $r= \vert \textbf{x}-\textbf{x}^\prime\vert $ and $\Pc_\zeta$ is the near scale invariant part of the power spectrum $P_\zeta$:
\begin{equation}
		P_\zeta  = \frac{H^2}{8\pi \epsilon} \approx A_s \LF \frac{k}{k_\ast} \RF^{n_s-1}\,. 
		\label{eq:HZpowe}
	\end{equation}
    where $n_s = 1-2\epsilon-\eta$. 
In standard inflation (SI) calculations, only the term $+$ (i.e., the wave with $ e^{-ik\tau}$ in Eq.\ref{modeqds}) is considered, and the slow-roll corrections are usually neglected. This resulting in the (near) scale-invariant power spectrum {(See Appendix G in \cite{Gaztanaga:2024vtr} for more details)}. {In DSI, the correction, $\Delta \Pc_v$ comes with positive and negative contributions to the (near) scalar-invariant part Eq.~\ref{eq:HZpowe} at parity conjugate points because we apply (quantum mechanically) the time reversality Eq.~\ref{dsisym} to the quasi-de Sitter vacuums $\vert 0_\pm\rangle_{\rm qdS}$ Eq.~\ref{qdSmat} at the parity conjugate regions.  } 
	
From the latest Planck data \cite{Planck:2018jri}  $A_s=2.2\times 10^{-9}$ and the value of spectral index $n_s$ corresponding to the pivot scale $k_\ast =a_\ast H_\ast = 0.05 {\rm Mpc}^{-1}$ is 
	\begin{equation}
		n_s = 1-2\epsilon_\ast-\eta_\ast \approx  0.9634 \pm 0.0048\quad \LF \rm{Planck TT+TE+EE} \RF
	\end{equation}  
In DQFT inflation (DSI), we include both terms as a direct-sum, which results in Eq.\ref{eq:vpm} and leads to Eq.\ref{eq:pwdsi} with:
\begin{equation}
\begin{aligned}
		\Delta \Pc_v  = \LF 1-n_s \RF\operatorname{Re}\LT \frac{2}{H_{3/2}^{(1)} \LF  \frac{k}{k_\ast} \RF} \frac{\pd H^{(1)}_{\nu_s}\LF \frac{k}{k_\ast} \RF}{\pd\nu_s} \Bigg\vert_{\nu_s=\frac{3}{2}} \RT . 
		\label{eq:delpPR}
  \end{aligned}
	\end{equation}	
Note that the $\pm$ in Eq.~\ref{eq:pwdsi} comes from the difference in the quantum vacuum Eq.~\ref{qdSmat} in the parity conjugate regions emerging from the time reversal operation Eq.~\ref{dsisym}. In other words, the additional term $\Delta \Pc_v$  in Eq.~\ref{eq:pwdsi} corresponds to a purely anti-symmetric contribution that changes sign at parity conjugate points.
In spherical coordinates, the parity transformation maps a point at radial distance $r$ from angular position $(\theta,\,\varphi)$ to its antipode, $(\pi - \theta,\, \pi + \varphi)$. This transformation cannot be realized by any rotation, and thus represents a discrete global symmetry distinct from statistical isotropy. 
It is important to emphasize that this global spatial parity asymmetry is unrelated to the observed birefringence in the CMB polarization, which arises from parity-violating interactions with well-defined transformation properties~\cite{Komatsu:2022nvu}.

 \begin{figure}
    \centering \includegraphics[width=0.9\linewidth]{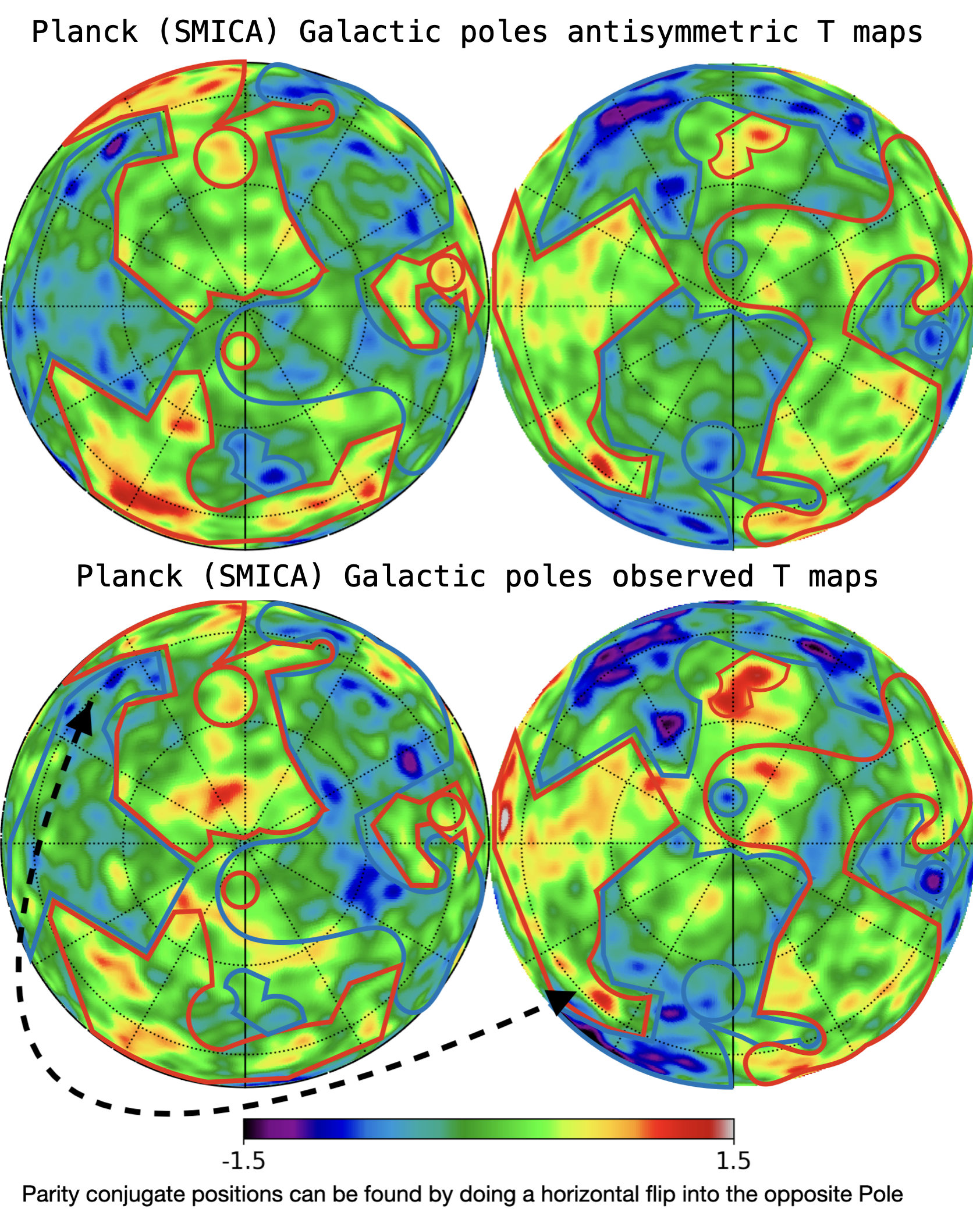}
    \caption{
    The top panels show the anti-symmetric, $\Tc^-(\hat{n}) = \LT\Ts(\hat{n})-\Ts(-\hat{n})\RT/2$, component of the Planck 2018 CMB temperature map (\cite{Akrami:2019bkn}) in North and South polar galactic caps. The visualization is such that the Galactic plane, shown as a grey mask in Fig.~\ref{fig:Zsmica},
    remains obscured along the edges. Parity conjugate positions can be identified by 
    horizontally flipping into the opposite pole (as indicated by the dashed arrow example) 
    where we find similar structures with opposing signs: $\Tc^-(-\hat{n})=-\Tc^-(\hat{n})$. The bottom panel presents the observed $\Tc=\Tc^+ + \Tc^-$ maps, including both the symmetric $\Tc^+(\hat{n}) = \LT\Ts(\hat{n})+\Ts(-\hat{n})\RT/2$ and anti-symmetric components. The evident resemblance between the top and bottom panels indicates the prevalence of the anti-symmetric component, in perfect agreement with DSI predictions Eq.\ref{eq:Cl_DQFT}.} 
    \label{fig:CMBcaps}
\end{figure}

\subsection{CMB temperature sky fluctuations}

On very large, super-horizon scales, the dominant contribution to CMB anisotropies arises from the Sachs-Wolfe effect \cite{1967ApJ...147...73S}, which corresponds to the gravitational redshift of the CMB photons due to small variations in curvature perturbations. This results in a one to one correspondence between temperature and metric fluctuations: $\Delta T\LF \hat{n} \RF  \sim \zeta$. The observed CMB temperature fluctuation map $\Ts\LF \hat{n} \RF \equiv \Delta T\LF \hat{n} \RF / T_0$ over the sky position $\hat{n}$ can be decomposed into spherical harmonics:
\begin{equation}
	\Ts\LF \hat{n} \RF	 = \sum a_{\ell m} Y_{\ell m}\LF \hat{n} \RF,
		 \quad a_{\ell m} = \int \sin\theta \ d\theta \ d\varphi \ \Ts(\hat n) \ Y_{\ell m}^\ast\LF \theta,\,\varphi \RF ; \quad C_{\ell} =  \sum_m \frac{\vert a_{\ell m}\vert^2}{2\ell+1}  
         \label{eq:alm}
\end{equation}
For a Gaussian field, the statistical properties of the Fourier-like harmonic coefficients $a_{\ell m}$ are characterized by the angular power spectrum $C_{\ell}$, as defined above. 
For antipodal (parity-conjugate) sky positions $(-\hat{n})$, the spherical harmonics $Y_{\ell m}$ satisfy:
\begin{equation}
Y_{\ell m}(-\hat{n}) = Y_{\ell m}\LF \pi-\theta,\,\pi+\varphi \RF = \LF -1 \RF^\ell Y_{\ell m}\LF \theta,\,\varphi \RF = \LF -1 \RF^\ell  Y_{\ell m}(\hat{n}). 
\label{eq:Tparity}
\end{equation}
The additional DSI term $\pm\Delta v$ in Eq.~\ref{eq:vpm} corresponds to a purely antisymmetric contribution that changes sign at parity-conjugate points (due to time-reversal operation). This translates into an additional purely antisymmetric temperature fluctuation $\delta T$ in the DSI model superimposed to the standard inflation (SI) contribution $\Ts^{SI}(\hat n)$:
\begin{equation}
     \Ts^{DSI}\LF \hat n \RF =  \Ts^{SI}(\hat n) \LF 1+\delta\Ts \RF \quad ; \quad \delta\Ts\LF \hat n \RF = -\delta \Ts\LF -\hat n \RF\,  \implies
     \delta\Ts\LF \hat n \RF = \sum_{\ell,\,m} \LF -1 \RF^{\ell+1} \delta a_{\ell m} Y_{\ell m} \LF \theta,\,\varphi \RF 
     \label{tempdecomp}
\end{equation}
This shows that when the fluctuations $\delta a_{\ell m}$ are smoothed and do not vary significantly from one multipole to the next, the additional antisymmetric DSI contribution results in an alternating negative (for even multipoles) and positive (for odd multipoles) modulation on top of the SI signal.

In the Sachs-Wolfe regime, we can relate the 2D angular power spectrum $C_{\ell}$ to the metric 3D metric power spectrum $P_\zeta(k)$ by a simple projection (see e.g. \cite{Durrer:2020fza}):
\begin{equation}
    C_{\ell}  = \frac{2}{9\pi} \int \frac{dk}{k} \Pc_\zeta\LF k \RF j_{\ell}^2\LF k/k_s \RF \,
    \label{eq:Cllstan}
\end{equation}
where $k_s=7\times 10^{-5} \,{\rm Mpc}^{-1}$ and $j_{\ell}\LF z \RF$ are spherical Bessel functions of the first kind.
Because the time-reversal symmetry $\Tc$ is broken by the slow-roll parameters in Eq.~\ref{dsisym}, we expect the quantum field during inflation to evolve asymmetrically in time at parity-conjugate points in physical space (resulting in Eq.\ref{eq:vpm} and Eq.~\ref{eq:pwdsi}). Following Eq.~\ref{eq:pwdsi} and Eq.~\ref{tempdecomp}, we obtain an angular power spectrum $C_{\ell}$ of the CMB that exhibits an excess of power in odd multipoles compared to even ones, as described below in Eq.\ref{eq:Cl_DQFT}.

In Fig.~\ref{fig:CMBcaps}, we can visually identify this parity asymmetry by the strong resemblance between the odd-parity component of the temperature map and the original CMB maps from Planck data \cite{Akrami:2019bkn}. This reflects that the CMB temperature maps contain an additional pure odd-parity component, as indicated by Eq.~\ref{tempdecomp}. The DSI framework naturally generates this additional odd symmetry.

It is worth noting that if we take the average of even and odd power spectra, we recover the well-known near scale-invariant power spectrum of standard inflation (SI) with $\Delta \Pc_v \simeq 0$. The advantage of the DSI model is that it requires no additional free parameters: the parity asymmetry depends solely on the spectral index $n_s \approx 0.9634 \pm 0.0048$ at the pivot scale $k_\ast = 0.05\, {\rm Mpc}^{-1}$ ($\ell\approx 800$), as measured by Planck \cite{Planck:2018jri}. {Inflationary quantum fluctuations in DSI are non-markovian (See Sec.~5.4 of \cite{Gaztanaga:2024vtr}). So compute here the scale-dependent features of power spectra for the large wavelength modes that affect the low-$\ell$ features of the CMB angular power spectrum. }

\subsection{DSI vs SI Model Comparison and Simulations}

To assess the significance of the observed parity asymmetry in the CMB, we compare $10^6$ simulated realizations of the data under two models: the SI model and the DSI model. Specifically, we evaluate the posterior probability $p(M|D)$ of each model $M$ given the data $D$.

This approach contrasts with the standard practice in the CMB community, which typically estimates the likelihood $p(D|M)$—the probability of the data given a specific model—to assess the significance of low multipole anomalies. However, this likelihood-based method presupposes the model and can lead to inflated uncertainties, especially because the $\Lambda$CDM model with SI generally predicts more large-scale power than is observed. This mismatch increases the sampling variance, thereby reducing the apparent significance of observed parity anomalies. For example, the low measured quadrupole $C_2$ has $p(D|M)=2.62\%$ while the posterior value is 29 times smaller: $p(M|D)=0.09\%$, as shown in Table \ref{tab1:annomalies}.

\begin{figure}
    \centering
    \includegraphics[width=0.6\linewidth]{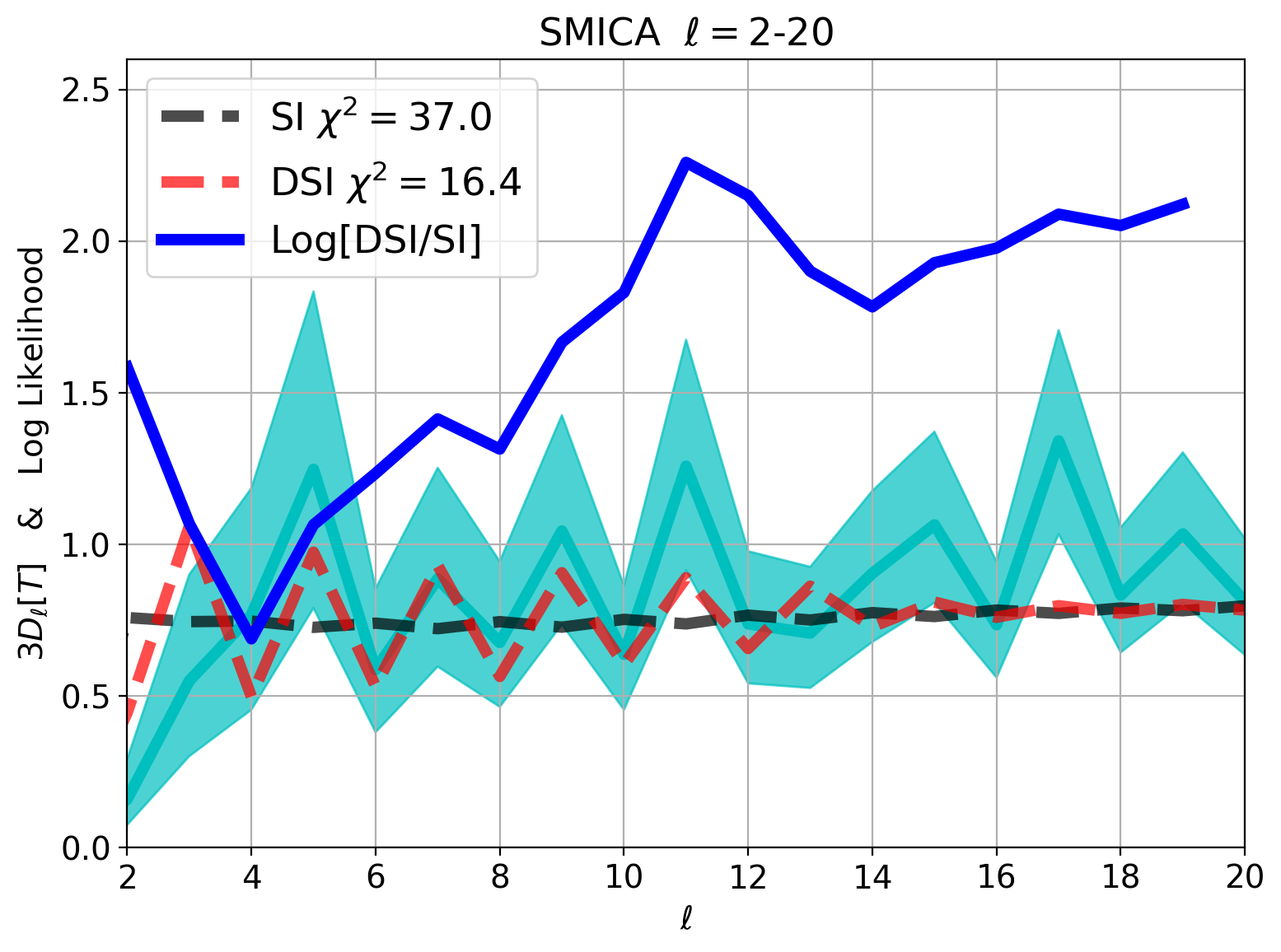}
    \includegraphics[width=0.6\linewidth]{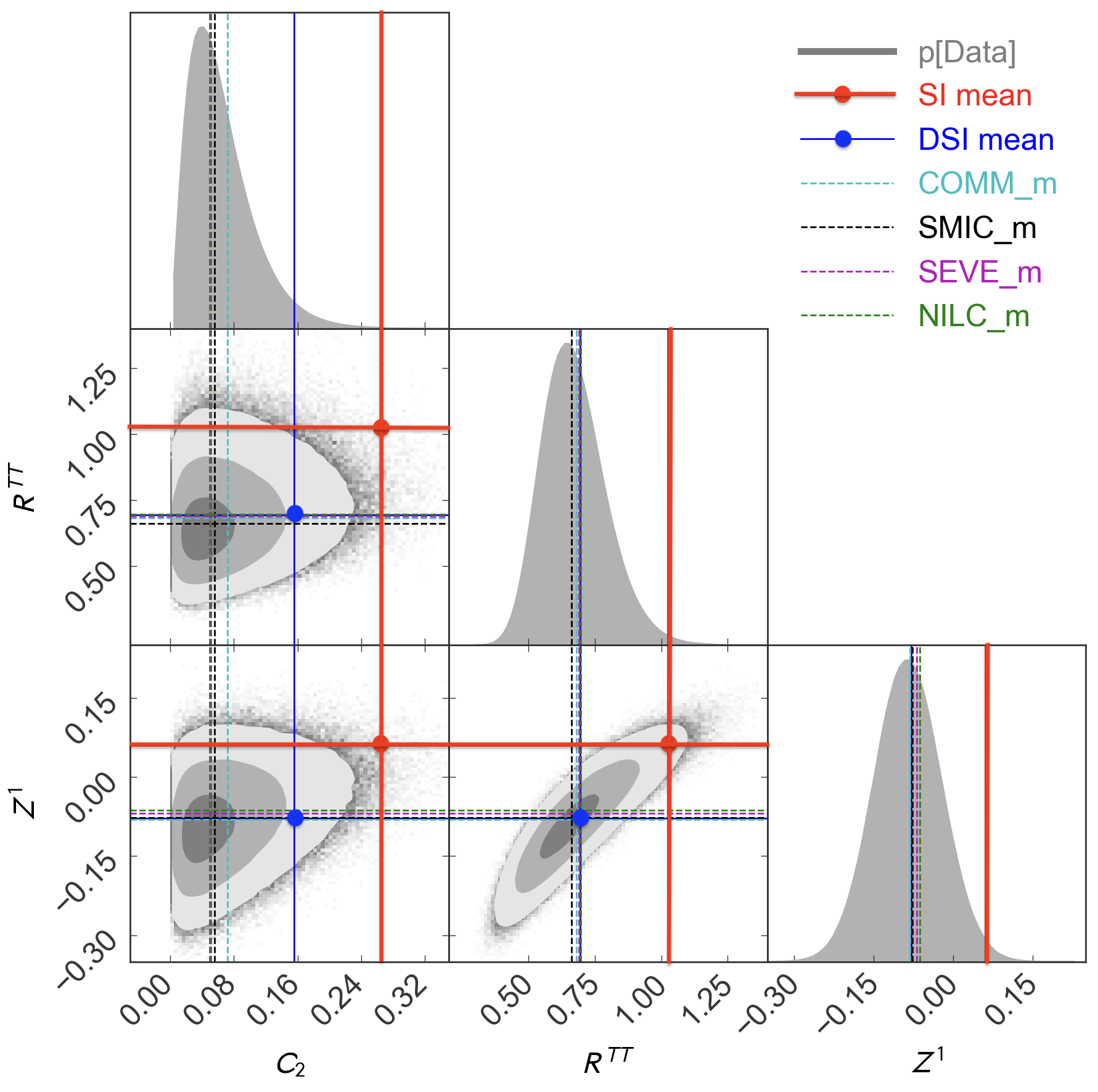} 
    \caption{\textbf{TOP:} Mean measured power spectrum $C_\ell$ (shown as $D_\ell \equiv \ell(\ell+1)\, C_\ell / 2\pi$, normalized such that the map has unit pixel variance) from the Planck 2018 SMICA$_m$ CMB temperature map (cyan line). The shaded cyan band shows the 68\% spread from $10^6$ realizations of the mean $C_\ell$.
    The black dashed line corresponds to the Standard Inflation (SI) model, i.e., the best-fit scale-invariant $\Lambda$CDM model primarily constrained by the high-$\ell$ range ($30 \leq \ell \leq 2500$). The red dashed line shows the prediction from the Direct-Sum Inflation (DSI) model defined in Eq.~\ref{eq:Cl_DQFT}, with no additional free parameters.
    The blue line at the bottom shows the Bayes fator: logarithm of the ratio of cumulative posterior likelihoods, $\log \left[ P_{\mathrm{DSI}}(<\ell) / P_{\mathrm{SI}}(<\ell) \right]$, as a function of $\ell$, demonstrating that the DSI model is up to 150 times more likely than SI on large angular scales ($\ell < 20$).
    \textbf{BOTTOM:} Parity asymmetry measurements derived from CMB temperature map realizations. The grey shaded regions indicate 1, 2, and 3$\sigma$ contours from simulations. Predictions from the SI and DSI models are shown in red and blue, respectively. The DSI model is over 650 times more likely than SI, based on full-map realizations.
    These simulation-based likelihood estimates are more reliable than those based on the $\chi^2$ of $C_\ell$ in the top panel, since the latter assume uncorrelated Gaussian likelihoods—a poor approximation for $C_\ell$, though more accurate for map-based estimators. 
    Dotted horizontal lines show mean estimates from different Planck component-separated maps and masks, demonstrating that systematic uncertainties are small compared to the statistical fluctuations in the $C_\ell$ realizations.}
    \label{fig:allpdf}
\end{figure}

Both the data and simulations are processed in the same way. Full details of the simulation setup are provided in Appendix D of \cite{Gaztanaga:2024vtr}.  Here we focus on the comparison of the $10^6$ all-sky-masked realizations of the measured $C_\ell$ power spectrum. We used the $C_\ell$ values measured by the Planck 2018 analysis.  We consider four Planck component-separated map estimates: {\sc Commander} (COMM$_m$), {\sc SMICA} (SMIC$_m$), {\sc SEVEM} (SEVE$_m$), and {\sc NILC} (NILC$_m$), each analyzed at $N_{\rm side} = 2048$. The corresponding sky fractions are $f_{\rm sky} = 88.8\%$, $84.2\%$, $83.8\%$, and $78.6\%$, respectively. The appropriate mask is applied to each simulation set. We use the \texttt{HEALPix} software package\footnote{\url{https://healpix.sourceforge.io/}} for all-sky map analysis and visualizations. The SI model corresponds to the Planck 2018 best fit scale invariant $\Lambda$ CDM angular power spectrum, denoted $C_\ell^{\rm SI}$ (shown as a black dashed line in the top panel of Fig.~\ref{fig:allpdf}) and goes throught the middle of the $C_\ell$ measurements (cyan line and shaded region 68\%), as expected.
The DSI model introduces a parity-violating modulation to the SI spectrum as follows:
\begin{equation}
    C_\ell^{\rm DSI} = C_\ell^{\rm SI} \left[1 + (-1)^{\ell+1} \Delta \mathcal{C}_\ell \right]
    \label{eq:Cl_DQFT}
\end{equation}
where the fractional modulation $\Delta \mathcal{C}_\ell$ is:
\begin{equation}
    \Delta \mathcal{C}_\ell = \frac{1}{C_\ell^{\rm SI}} \int_0^{k_c} \frac{dk}{k} A_s\left( \frac{k}{k_s} \right)^{n_s - 1} j_\ell^2\left( \frac{k}{k_s} \right) \Delta \mathcal{P}_v(k)
    \label{eq:RDcl}
\end{equation}
Here, $\Delta \mathcal{P}_v(k)$ is given in Eq.~\ref{eq:delpPR}, and the standard $\Lambda$CDM angular power spectrum is (see  Eq.\ref{eq:Cllstan}) :
\begin{equation}
    C_\ell^{\rm SI} = \int_0^{\infty} \frac{dk}{k} A_s\left( \frac{k}{k_\ast} \right)^{n_s - 1} j_\ell^2\left( \frac{k}{k_s} \right)
    \label{eq:SICl}
\end{equation}
The SI prediction corresponds to $
\Delta \Pc_v =0$, while the DSI one is given by Eq.\ref{eq:delpPR}.
 A cut-scale $k_c = 0.02 k_\ast$ is used in Eq.\ref{eq:RDcl} as Eq.\ref{eq:delpPR} is accurate enough for low-$\ell$ or large angular scales. This cutoff scale is also related to the coarse-graining scale of stochastic inflation (\cite{Gaztanaga:2024vtr}), which determines the large wavelength modes that have already become classical on the onset of inflation while the mode $k_\ast$ exits the horizon. In DSI, these large-wavelength modes create parity-asymmetric inhomogeneities that correct the FLRW spacetime on large scales. The parity asymmetry ceases to exist towards the small scales because the $\Pc\Tc$ symmetry is recovered towards the short-distance scales. This can be seen in Fig.~11 of \cite{Gaztanaga:2024vtr}. An intuitive explanation for this is that large wavelength modes experience quantum gravitational effects more than those of short wavelength modes. For the small scales $k>k_c$, the predictions of DSI become almost the same as with the (near) scale-invariant power spectrum as contributions from $\Delta \Pc_v$ term become negligible.

The resulting modulation in Eq.~\ref{eq:Cl_DQFT} oscillates with amplitude $\sim 20\%$ for multipoles $\ell < 10$, producing a measurable parity asymmetry (see dashed red line in top panel of Fig.~\ref{fig:allpdf}). We generate $10^6$ Monte Carlo realizations for both SI and DSI models, allowing a direct and statistically robust comparison with the Planck data.
 
 In Fig.~\ref{fig:allpdf} and Table \ref{tab1:annomalies} we present how the posterior probabilities from the $10^6$ realizations of the observational data very significantly prefer the mean DSI model prediction over the SI model. We compare the quadrupole amplitude $C_2=C_{\ell=2}$, the even-to-odd multipole ratio of angular power spectra:
 \begin{equation}
 R^{TT} = \frac{\sum_{\ell=even}^{\ell max=20} \ell (\ell+1) C_{\ell}}{\sum_{\ell=odd}^{\ell max=20} \ell (\ell+1) C_{\ell}}     
 \end{equation}
\
\begin{table}
		\caption{Parity posterior probability  $p[M|D]$ ($\times10^2$, i.e. in $\%$) of a model  given the data (based on $10^6$ sky realizations of the data). Each line corresponds to a different CMB parity indicator. We compared two different models: Standard  Inflation quantum fluctuations with LCDM (SI) and direct-sum inflationary quantum fluctuations (DSI). 
			Data is estimated from the Planck 2018 masked SMICA component separation map. Very similar results are found for the other maps (see dashed lines in Fig.\ref{fig:allpdf}).
			In the last column, 'ratio' refers to the ratio of posterior probabilities between the DSI and SI predictions. In the second column, we also show for reference the non-posterior likelihood $p[D|M]$.}
		\begin{center}
			\label{tab1:annomalies}
			\begin{tabular}{l | c | c  c | c}
				Parity  & SI  & SI & DSI  & ratio  \\  indicator & \ $p[D|M]$ & \ $p[M|D]$ \ & \ $p[M|D]$ \ &  \  DSI/SI \\ \hline  \hline 
				$C_2$ & 2.62\ \% & 0.09\ \% & 3.3\ \% & 37 \\         
				$R^{TT}$ & 1.0\ \%  & 0.7\ \% & 39.5\ \% & 56  \\ 
				$Z^1$ & 3.89\ \%  & 1.12\ \% & 45.3\ \% & 40 \\            
				$C_2 ,R^{TT}$  & 0.12\ \% &  0.003\ \% &  1.96\ \% & 653 \\ 
				$Z^1 ,R^{TT}$  & &  0.45\ \% &  34.6\ \% & 77 \\
				$Z^1 , C_2$  &  &  0.016\ \% &   2.65\ \% & 166 \\ \hline
				\hline
			\end{tabular}
		\end{center}
	\end{table}

\begin{figure}
\centering
\includegraphics[width=0.9\linewidth]{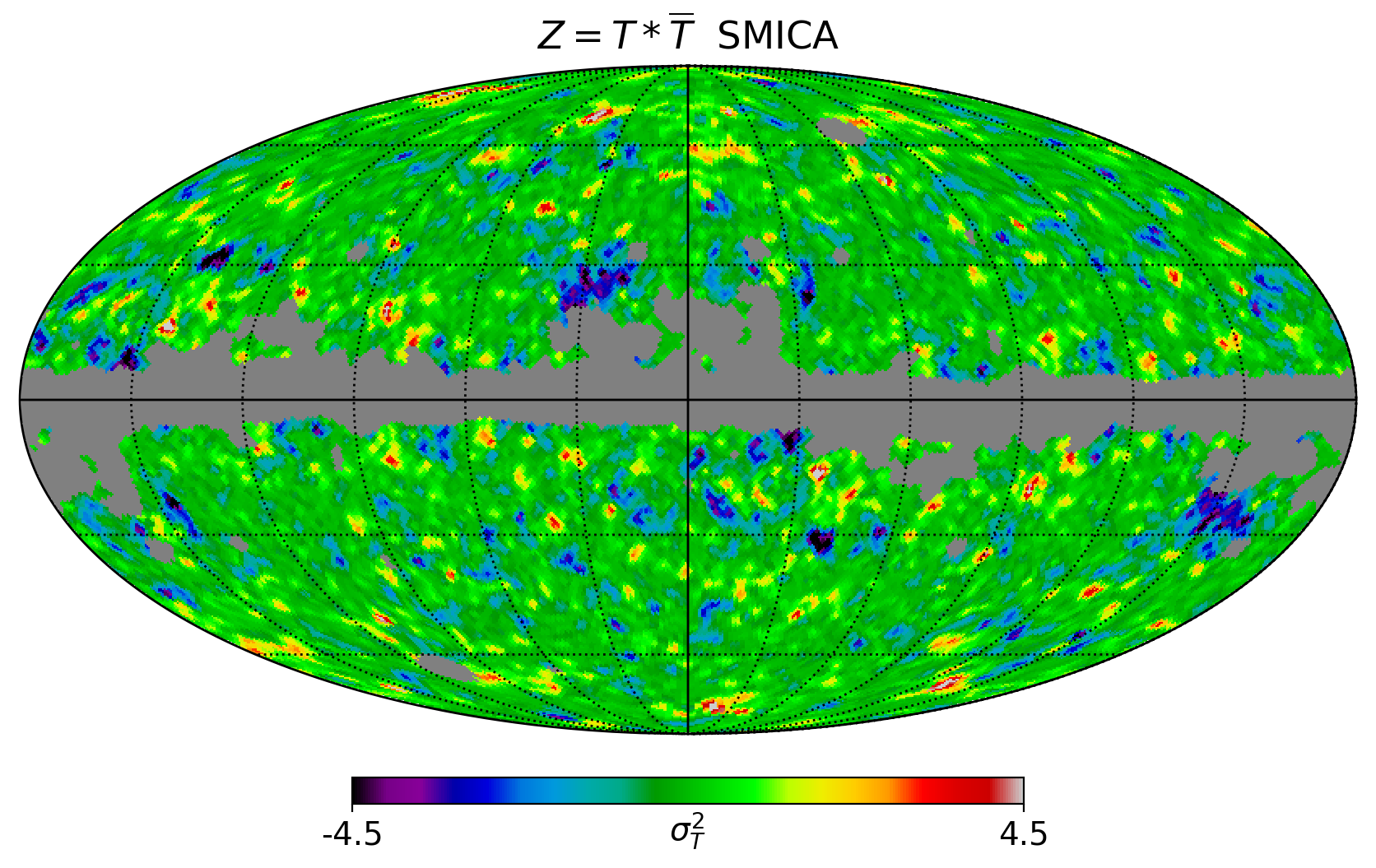}
\includegraphics[width=0.8\linewidth]{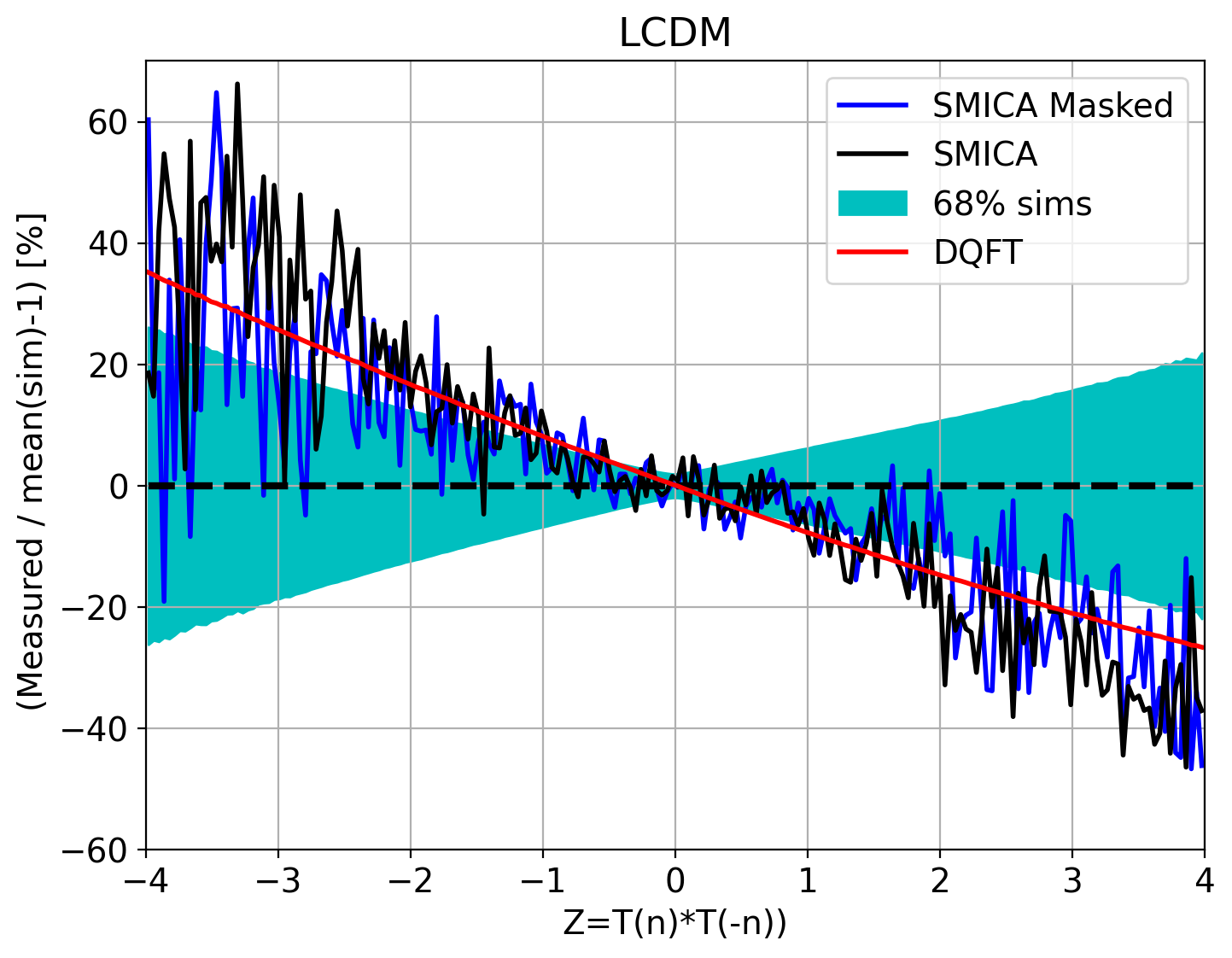}
\caption{
\textbf{TOP:} Parity map $Z(\hat{n}) = \mathcal{T}(\hat{n})\, \mathcal{T}(-\hat{n})$ derived from the SMICA map shown in Fig.~\ref{fig:CMBcaps}, including the Planck galactic mask. This map is invariant under the parity operator $\mathcal{P}$, and encodes the degree of symmetry between temperature values at antipodal sky positions. Positive regions (red) indicate even-parity (symmetric) structure in the CMB temperature field $\mathcal{T}$, while negative regions (blue) highlight odd-parity (antisymmetric) structure. The dominance of negative values visually indicates a preference for odd parity in the observed sky. Notably, the average value $\langle Z \rangle \simeq -0.1$ supports the DSI prediction, whereas the SI model predicts $\langle Z \rangle \simeq 0.1$ (see also Fig.~\ref{fig:allpdf}).
\textbf{BOTTOM:} Relative deviation (in percent) between the observed distribution of $Z(\hat{n})$ values (for both unmasked SMICA, black, and masked SMICA, red) and the mean SI prediction (black dashed line). The data show a clear excess of negative $Z$ values and a deficit of positive ones, compared to expectations under SI. The shaded cyan band represents the 68\% confidence interval from $10^6$ SI simulations. The red line shows the DSI model (labeled DQFT), which offers a significantly better match to the observed distribution. Together with Fig.~\ref{fig:CMBcaps}, this provides visual and statistical evidence for the underlying parity-violating quantum field theory in curved spacetime (QFTCS) responsible for the observed asymmetry.
}
\label{fig:Zsmica}
\end{figure}

\section{Conclusions}
\label{sec:concl}
In this paper, we underscore the pivotal role of unitary Quantum Field Theory in Curved Spacetime (QFTCS) in elucidating the intricacies of early Universe physics. Contrary to prevailing notions, we contend that the issue of unitarity in QFTCS is not inherently tied to quantum gravity at Planck scales, but rather stems from 
throwing away the vital role of $\Pc\Tc$ transformations {in the construction of vacuum} for quantum fields in curved spacetime. {In the context of de Sitter spacetime, which is $\Pc\Tc$ symmetric manifold}, {we illustrated that one can restore unitarity of QFTCS by taking into consideration geometrical aspects of spacetime.} 

With the $\Pc\Tc$ {based} formulation of the Schr\"odinger equation Eq.~\ref{eq:Schr}, {we embrace a new understanding of quantum theory where a single quantum state is expressed as direct-sum of two components that describe the parity conjugate regions of physical space with opposite arrows of time.}
{The DQFT framework constructs quantum fields with opposite arrows of time at parity conjugate regions of physical space. This new geometric construction of quantum fields echoes with discrete (a)symmetries of the spacetime manifold. Within this new description, the spacetime horizon acts as a $\Pc\Tc$ mirror which contains any information within the horizon in the form of pure states.} {This reinstates unitarity in de Sitter spacetime, which was previously thought to be lost in standard frameworks of quantization.} Fig.~\ref{fig:unitarity}--\ref{fig:NPD} 
visually represent this concept, illustrating how a hypothetical classical observer consistently experiences pure states without encountering information loss beyond the horizon. {The DQFT framework can be applied to any curved spacetime without any necessity of $\Pc\Tc$ symmetry in the manifold. The construction of DQFT only prescribes that every quantum field should be expressed via two components describing parity conjugate regions of physical space with opposite arrows of time, where the quantum vacuum is a geometrical direct-sum based on $\Pc\Tc$ operations. With this construction, quantum fields could evolve asymmetrically at parity-conjugate points if the time reversal symmetry is broken. Thus, in manifolds like de Sitter spacetime, where the manifold is $\Pc\Tc$ symmetric, quantum fields geometrically obey the symmetry and lead to equal two-point correlations at parity conjugate regions of physical space. }

Applying the framework of unitary QFTCS to inflationary quantum fluctuations, we derived a prediction for a parity asymmetry in the CMB originating from time-reversal symmetry breaking during inflation. {It is well-known that time is a parameter in quantum theory. If one considers quantum fluctuations in gravitational physics, all the time-dependent quantities that describe the dynamics of spacetime act as parameters. Thus, in quantum gravity (even at the linearized level, which is the case of inflationary quantum fluctuations), time reversibility is much more complex than a naive thinking of time coordinate reflection. 
In our construction, the slow-roll parameters that describe inflationary spacetime dynamics break the $\Pc\Tc$ symmetry of the de Sitter vacuum in DQFT, leading to parity asymmetric evolution of the inflationary quantum fluctuations. 
We have shown that this leads to non-trivial correlation between anti-podal points ($\LF \theta,\,\varphi \RF$ and $\LF \pi-\theta,\,\pi+\varphi\RF $) of the CMB sky.  }
The striking correlation illustrated in Fig.~\ref{fig:CMBcaps}--\ref{fig:Zsmica} between the observed hot and cold mirrored structures of the CMB map at antipodal points provides strong support for the unitary treatment of inflationary fluctuations within the DQFT framework.

Our findings indicate that superhorizon fluctuations at antipodes manifest as $\Pc\Tc$ conjugates, a prediction derived from direct-sum QFTCS (i.e., the direct-sum inflation, DSI). This prediction is strongly supported by the data, with a posterior probability over 650 times greater than that of standard inflation, which is based on non-unitary QFTCS (see Table~\ref{tab1:annomalies}).
The observed parity asymmetry also provides a unified explanation for other large-scale CMB anomalies, such as the so-called hemispherical power asymmetry \cite{Gaztanaga:2024vtr}.

To assess the significance of the low-multipole anomalies, we evaluate the posterior probability $p(M|D)$ of each model $M$ given the data $D$. This Bayesian approach contrasts with the standard practice in the CMB community, which evaluates $p(D|M)$—the likelihood of the data under the SI model. 
For example, the low measured quadrupole $C_2$ has $p(D|M)=2.62\%$ while the posterior value is 29 times smaller: $p(M|D)=0.09\%$, as shown in Table \ref{tab1:annomalies}.
The former approach tends to inflate uncertainties due to the excess power predicted by $\Lambda$CDM at large scales, which increases sampling variance and thereby reduces the apparent significance of observed parity asymmetry.  

{Our study highlights the importance of QFTCS and its non-trivial role in underpinning the quantum aspects of gravity. By sticking to the foundational aspects of QFTCS, we addressed the long-standing issues associated with the understanding of CMB anomalies. Our grounded approach not only shed new light on understanding quantum gravity but also allowed us to derive predictions that necessarily addressed the observations. This approach complements and poses new challenges to the past investigations \cite{Ashtekar:2021izi,Kitazawa:2014mca} that addressed CMB anomalies as evidence of yet unknown frameworks of Planck scale quantum gravity. }

\begin{acknowledgments}
EG acknowledges grants from the Spanish Plan Nacional (PGC2018-102021-B-100) and
Maria de Maeztu (CEX2020-001058-M).  KSK acknowledges the Royal Society for the Newton International Fellowship. 
\end{acknowledgments}

\bibliographystyle{apsrev4-2}
\bibliography{Anoqft.bib}

\appendix 

\section{Quantum harmonic oscillator in direct-sum quantum mechanics} 
\label{app:QHO}
The purpose of this section is to explicitly illustrate the application of direct-sum Schr\"{o}dinger to the context of the simple harmonic oscillator whose Hamiltonian is given by 
\begin{equation}
     \hat{\mathbb{H}} = \frac{E}{2} \LF \hat p^2 + \hat x^2 \RF = \underbrace{\frac{E}{2} \LF \hat p_+^2 + \hat x_+^2 \RF}_{\hat{\mathbb{H}}_+} \oplus \underbrace{\frac{E}{2} \LF \hat p_-^2 + \hat x_-^2 \RF}_{\hat{\mathbb{H}}_-} = \begin{pmatrix}
         \hat{\mathbb{H}}_+ && 0 \\ 
         0 && \hat{\mathbb{H}}_-
     \end{pmatrix}
     \label{HOhami}
\end{equation}  
Following Eq.~\ref{disumwvf}, we express the operators $\hat x_\pm$ and $\hat p_\pm$ in terms of the two pairs of creation and annihilation operators $\LF a,\, a^\dagger \RF$ and $\LF b,\,b^\dagger \RF$ associated with the sectorial Hilbert spaces $
\Hc_\pm$ as
\begin{equation}
		\begin{aligned}
			\hat{x}_+ & = \sqrt{\frac{1}{2}}\LF a+a^\dagger \RF,\quad \hat{p}_+ = -i\frac{d}{dx_+}= \frac{i}{\sqrt{2}} \LF a^\dagger-a \RF \\ 
			\hat{x}_- & = \sqrt{\frac{1}{2}}\LF b+b^\dagger \RF,\quad \hat{p}_- = i\frac{d}{dx_-}= -\frac{i}{\sqrt{2}} \LF b^\dagger-b \RF 
		\end{aligned}
	\end{equation}
	with 
	\begin{equation}
		\begin{aligned}
			\Big[\hat{x}_+,\,\hat{p}_+\Big] & = i,\quad \Big[\hat{x}_-,\,\hat{p}_-\Big]= -i\\
			\Big[a,\,a^\dagger\Big] & =     \Big[b,\,b^\dagger\Big] =1,\quad  \Big[a,\,b^\dagger\Big]=  \Big[a,\,b\Big]=0\,. 
		\end{aligned}
	\end{equation}

\begin{figure}
    \centering
    \includegraphics[width=0.5\linewidth]{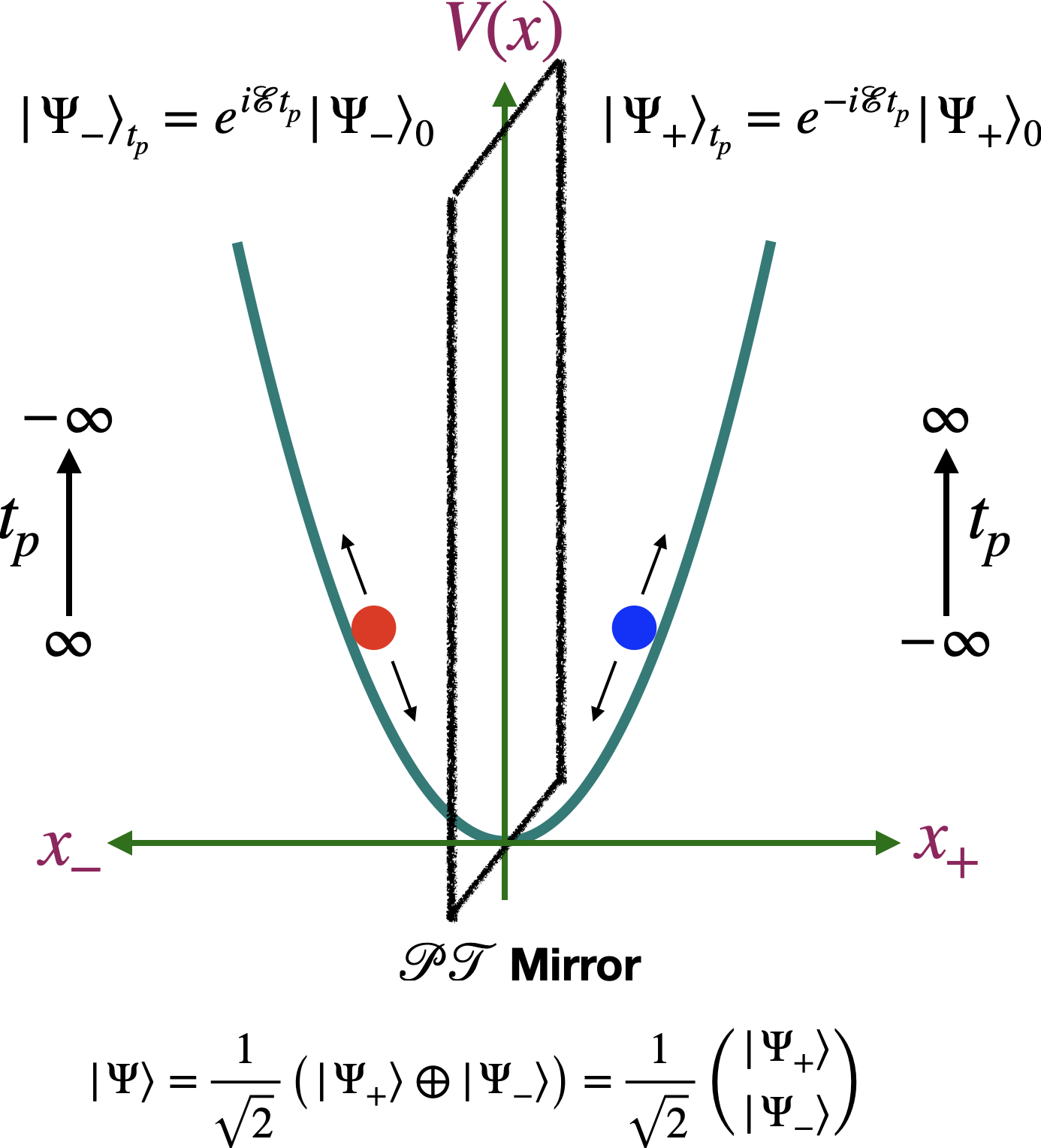}
    \caption{This represents the schematic understanding of the quantum harmonic oscillator in direct-sum quantum mechanics in which a single quantum state is written as direct-sum of two components $\vert \Psi\rangle = \frac{1}{\sqrt{2}}\LF \vert \Psi_+\rangle \oplus \vert \Psi_-\rangle \RF$ described by geometric superselection sector Hilbert spaces $\Hc_\pm$ (the total Hilbert space that represents the state $\vert \Psi\rangle$  that describe the parity conjugate regions of physical space where arrow of times are oppositely defined. Both $\vert\Psi_\pm \rangle$ are the positive energy states in the respective sectorial Hilbert spaces. }
    \label{fig:Hoshematic}
\end{figure}

Solving the direct-sum Schr\"{o}dinger equation Eq.~\ref{eq:Schr} for the direct-sum split harmonic oscillator Hamiltonian operator Eq.~\ref{HOhami} yields the wave function $\Psi_n\LF x \RF$ for the different energies $E_n = \frac{1}{2}\omega\LF n+\frac{1}{2} \RF$ as
\begin{equation}
   \Psi_n(x) = \langle x \vert \Psi_n\rangle \equiv \begin{cases}
       \frac{1}{\sqrt{2^{n+1}n!}}\LF \frac{1}{\pi} \RF^{1/4} e^{-\frac{1}{2}x_+^2} H_n\LF x_+ \RF e^{-iE_n t_p},\quad x_+\gtrsim 0\\ 
      \frac{1}{\sqrt{2^{n+1}n!}}\LF \frac{1}{\pi} \RF^{1/4} e^{-\frac{1}{2}x_-^2} H_n\LF x_- \RF e^{iE_n t_p},\quad x_- \lesssim 0
   \end{cases}
\end{equation}
where we followed Eq.~\ref{eq:wavefunction}. It is important to recall the description of quantum mechanics with two components $\vert\Psi\rangle $ of the parity conjugate regions with geometric superselection sectors $\Hc_\pm$, which reinstate the $\Pc\Tc$ symmetry of the physical system respected at the quantum mechanical level. In Fig.~\ref{fig:Hoshematic} we present the schematic description of the quantum harmonic oscillator in direct-sum QM. 

\section{Klein-Gordon field operator in DQFT}
\label{ap:KG}
The Klein-Gordon field operator for DQFT in Minkowski spacetime $ds^2 = -dt_m^2+d\textbf{x}^2$ split into direct-sum 
	\begin{equation}
			\hat{\phi}  = \frac{1}{2}  \hat{\phi}_{+}  \LF t_m,\, \textbf{x} \RF \oplus  \hat{\phi}_{-} \LF -t_m,\,-\textbf{x} \RF  
			 = \frac{1}{\sqrt{2}} \begin{pmatrix}
				\hat{\phi}_{+} & 0 \\ 
				0 & 	\hat{\phi}_{-}
			\end{pmatrix}
		\label{disum}
	\end{equation}
	where
	\begin{equation}
			\hat{\phi}_{\pm}   = 	\int \frac{d^3k}{\LF 2\pi\RF ^{3/2}}	\frac{1}{\sqrt{2\vert k_0 \vert}} \Bigg[\hat a_{(\pm)\,\textbf{k}}  e^{\pm ik\cdot x}+\hat a^\dagger_{(\pm)\,\textbf{k}} e^{\mp ik\cdot x} \Bigg]  
		\label{fiedDQFTMin}
	\end{equation}
 where $k\cdot x = -k_0t_m+\textbf{k}\cdot \textbf{x}$. 
	The creation and annihilation operators satisfy the canonical relations
	\begin{equation}
		\begin{aligned}
			[\hat{a}_{(\pm)\,\textbf{k}},\,\hat{a}_{(\pm)\,\textbf{k}^\prime}^\dagger]  & = \delta^{(3)}\LF \textbf{k}-\textbf{k}^\prime \RF,\quad [\hat{a}_{(\pm)\,\textbf{k}},\,\hat{a}_{(\mp)\,\textbf{k}^\prime}]=0 \\
            [\hat{a}_{(\pm)\,\textbf{k}},\,\hat{a}^\dagger_{(\mp)\,\textbf{k}^\prime}]& =0
		\end{aligned}
		\label{comcan}
	\end{equation}
 which defines the positive norm states in the direct-sum Minkowski vacuums 
 \begin{equation}
 \vert 0_M\rangle = \vert 0_{M_+}\rangle \oplus \vert 0_{M_-}\rangle =\begin{pmatrix}
     \vert 0_{M+}\rangle \\ \vert 0_{M_-}\rangle \end{pmatrix},\quad a_{(\pm)}\vert 0_{\pm}\rangle =0. 
 \label{minvac}
 \end{equation}
 The second condition in Eq.\ref{comcan} implies commutativity of operators $\LT  \hat\phi_+,\,\hat\phi_- \RT =0$, which can be seen as a new causality condition. 
 The
The Fock space in DQFT is the direct-sum of two Fock spaces (geometric superselection sectors) $\Fc_T = \Fc_+ \oplus \Fc_-$ corresponding to the components of field operators $\hat\phi_\pm$ that lead to the quantum states forward and backward in time at parity conjugate regions of position space. These field operators are defined through the pairs of creation and annihilation operators Eq.\ref{comcan}. Since the Minkowski spacetime is $\Pc\Tc$ symmetric, the scalar field operator in DQFT respects the same symmetry. Generalizing this structure to all quantum fields in Minkowski spacetime, such as complex scalar, vector, and fermionic degrees of freedom, is very straightforward. Any standard model (single particle) state, whether it is particle ($\vert \rm SM\rangle) $ or anti-particle ($\vert \overline{\rm SM}\rangle $, is expressed in DQFT as direct-sum of two components
\begin{equation}
    \vert \rm SM\rangle = \begin{pmatrix}
        \vert {\rm SM}_+\rangle \\ 
        \vert {\rm SM}_-\rangle 
    \end{pmatrix},\quad  \vert  \overline{  {\rm SM}} \rangle = \begin{pmatrix}
         \vert \overline{{\rm SM}}_+ \rangle  \\ 
         \vert \overline{{\rm SM}}_- \rangle 
    \end{pmatrix}
\end{equation}
with respect to $\Pc\Tc$ conjugate splitting of the vacuum $\vert 0_M\rangle  = \vert 0_{M_+}\rangle \oplus \vert 0_{M_-}\rangle $, {where the subscript $M_\pm$ mean $\Pc\Tc$ conjugate sheets of Minkowski spacetime}. {Also, we assume the same structure of the geometric superselection rule for all Fock spaces of
The SM degrees of freedom i.e., $\Pc\Tc$ conjugate regions of spacetime defined the same way for all quantum fields of SM.} Note that in each vacuum $\vert 0_{M\pm}\rangle$ a (component) of an anti-particle state can be interpreted as a particle (component) state going backward in time. It means, $\vert \overline{{\rm SM}}_+ \rangle$ can be interpreted as $\vert {\rm SM}_+\rangle$ state going backward in time with respect to vacuum $\vert 0_{M+}\rangle$. In the similar way, $\vert \overline{{\rm SM}}_- \rangle$ can be interpreted as $\vert {\rm SM}_-\rangle$ state going backward in time with respect to vacuum $\vert 0_{M-}\rangle$. Thus, Feynman's interpretation remains the same. Remember that in DQFT, a single quantum state is expressed as direct-sum of two components related by $\Pc\Tc$ transformations.  
{To be explicit, the Fermionic field operator in DQFT is quantized as
   \begin{equation}
   \begin{aligned}
  \hat \psi & = \frac{1}{\sqrt{2}}\LF \hat \psi_+\oplus \hat \psi_- \RF \\
     \hat  \psi_{\pm} & = \sum_{{\tilde s}} \int \frac{d^3k}{\LF 2\pi \RF^{3/2}\sqrt{2\vert k_0\vert}} \Bigg[ c_{{\tilde s}(\pm)\textbf{k}} u_{\tilde s}(\textbf{k}) e^{\pm ik\cdot x} + d_{{\tilde s}(\pm)\textbf{k}}^\dagger v_{\tilde s}(\textbf{k}) e^{\mp ik\cdot x}\Bigg]
     \end{aligned}
   \end{equation}
where ${\tilde s}=1,2$ correspond to the two independent solutions of $\LF \slashed k+m\RF u_s=0$ and $\LF -\slashed k+m\RF v_s=0$ corresponding to spin-$\pm\frac{1}{2}$. The creation and annihilation operators of geometric superselection sectors here satisfy the anti-commutation relations $\Big\{ c_{s(\pm)\textbf{k}},\,c_{s(\pm)\textbf{k}}^\dagger \Big\}=1,\, \Big\{ c_{s(\mp)\textbf{k}},\,c_{s(\pm)\textbf{k}}^\dagger \Big\}=\Big\{ c_{s(\mp)\textbf{k}},\,c_{s(\pm)\textbf{k}} \Big\}=0$ leading to the new causality condition $\Big\{ \hat\psi_+,\,\hat \psi_-\Big\} =0$.} For further details on DQFT in particular, the CPT (charge conjugation, Parity and Time reversal operations)
invariance of scattering amplitudes, which holds in both $\Pc\Tc$ based geometric superselection sectors \eqref{minvac}, and quantizing other fields of the standard model, see \cite{Kumar:2023ctp}. {We caution that our $\Pc\Tc$ based quantization scheme which geometrically splits the quantum fields as function of space and time
is not to be confused with CPT theorem, CP violation aspects of standard model, which holds anyway in our formulation too. 
Once more, in DQFT we construct quantum fields operators respecting the $\Pc\Tc$ symmetry ($\textbf{x}\to -\textbf{x}$,\,$t_m\to -t_m$) of Miknkowski spacetime and it does not change any results of standard model because all the interaction terms are split into direct-sum structure in the following way 
\begin{equation}
   \Lc_c \sim\Oc_{SM}^3=\begin{pmatrix}
        \Oc_{SM_+}^3 & 0 \\ 
        0 & \Oc_{SM_-}^3
    \end{pmatrix} \quad \Lc_q \sim \Oc_{SM}^4 = \begin{pmatrix}
        \Oc_{SM_+}^4 & 0 \\ 
        0 & \Oc_{SM_-}^4
    \end{pmatrix}
\end{equation}
Here, $\Oc_{SM}$ is an arbitrary operator involving any standard model fields and their derivatives. Thus, CPT invariance and CP violations in weak interactions of standard model\footnote{The discrete symmetries such as CPT, CP in QFT and standard model describing scattering amplitudes and certain decay rates of heavy particles are associated with applying C, P, and T on the states of momentum space. To be precise, under charge conjugation $C$, we interchange charge, and under PT, the 3-momenta are unchanged, but the definition of ingoing state gets interchanged with the outgoing state. The combined operation of CPT turns ingoing particles into outgoing antiparticles and vice versa in a scattering process.} remains unchanged. This is because standard $C,\,P,\,T$ operations can be applied separately in $\vert 0_\pm\rangle_M$ mean charge conjugation, changing the 3-momenta $\textbf{k}\to \textbf{k}$. 
This means, once we split the quantum fields into geometric superselection sectors, we can apply discrete operations on standard model degrees of freedom as
\begin{equation}
    {\rm CPT}\vert \rm SM\rangle =  {\rm CPT}\vert \rm SM_+\rangle \oplus  {\rm CPT}\vert \rm SM_-\rangle 
\end{equation}.
 }

\section{DQFT in dS}
\label{ap:dS}

Applying DQFT in dS (\cite{Gaztanaga:2024vtr,Kumar:2023ctp}), the canonically rescaled scalar field\footnote{Here we mean some scalar field $\phi$, not to be confused with the inflaton field} ($\phi\to e^{Ht}\phi$) operator is 
\begin{equation}
		\begin{aligned}
			\hat{\phi} & = \frac{1}{\sqrt{2}}  \hat{\phi}_{+}  \LF \tau,\, \textbf{x} \RF \oplus \frac{1}{\sqrt{2}} \hat{\phi}_{-} \LF -\tau\,-\textbf{x} \RF \,,
		\end{aligned} = \begin{pmatrix}
		    \hat \phi_+\LF \tau,\, \textbf{x} \RF && 0 \\ 
            0 && \hat\phi_-\LF -\tau,\, -\textbf{x} \RF 
		\end{pmatrix}
		\label{disum-dS}
	\end{equation}
	where 
	\begin{equation}
		\begin{aligned}
			\hat{\phi}_{\pm}   &  = 	\int \frac{d^3k}{\LF 2\pi\RF ^{3/2}}	 \Bigg[\hat d_{(\pm)\,\textbf{k}} \phi_{\pm\,k}   e^{\pm i\textbf{k}\cdot \textbf{x}}+\hat d^\dagger_{(\pm)\,\textbf{k}} \phi_{\pm\,k}^\ast   e^{\mp i\textbf{k}\cdot \textbf{x}} \Bigg]  
		\end{aligned}
		\label{fiedDQFTdS}
	\end{equation}
Here the operators $d_{(\pm)\,\textbf{k}},\,d^\dagger_{(\pm)\,\textbf{k}}$ satisfy the similar relations like in Eq.\ref{comcan}. 
The dS mode functions are:
\begin{equation}
\phi_{\pm\,k} = \frac{1}{\sqrt{2k}} \LF 1\mp \frac{i}{k\tau} \RF e^{\mp ik\tau} = \frac{\sqrt{\mp\pi \tau}}{2} H_{3/2}^{(1)}\LF \mp k\tau \RF
\label{eq:dSmodes}
\end{equation}
corresponding to the Bunch-Davies vacuum of the DQFT
{\begin{equation}
    \vert 0\rangle_{\rm dS} = \begin{pmatrix}
           \vert 0_+\rangle_{\rm dS} \\ 
               \vert 0_-\rangle_{\rm dS} 
    \end{pmatrix}
\end{equation}}
which recovers short-distance behavior of quantum fields in Minkowski Eq.~\ref{fiedDQFTMin} {i.e., 
\begin{equation}
    \hat{\phi}_{\pm} \Big\vert_{k\tau\gg 1} = \frac{1}{\sqrt{2k}}e^{\mp ik\tau}
\end{equation}
The two point correlations in de Sitter spacetime are identical at parity conjugate regions 
\begin{equation}
	\begin{aligned}
		& \frac{1}{a^2}{}_{\rm dS}\langle 0_+\vert \hat{\phi}_{+}\LF \tau,\, \textbf{x} \RF \hat{\phi}_{+}\LF \tau,\, \textbf{x}^\prime \RF\vert 0_+\rangle_{\rm dS} = \\ & \frac{1}{a^2} {}_{dS}\langle 0_-\vert \hat{\phi}_-\LF -\tau,\, -\textbf{x} \RF \hat{\varphi}_{-}\LF -\tau,\, -\textbf{x}^\prime \RF\vert 0\rangle_{dS-} = \int \frac{dk}{k}\frac{\sin k\xi}{k\xi} \frac{H^2}{4\pi^2}\,, 
	\end{aligned}
	\label{eqcorr}
\end{equation}
where $\xi = \vert \textbf{x}-\textbf{x}^\prime\vert$. }

\end{document}